\documentclass[twocolumn]{aastex62}
\usepackage{natbib}
\graphicspath{{./}{pictures/}}

\received{}
\revised{}
\accepted{}
\submitjournal{ApJ}

\shorttitle{Tidal disruption events}
\shortauthors{Ryu et al.}

\usepackage{graphicx}	
\usepackage{amsmath}	
\usepackage{amssymb}	
\usepackage{cases}
\usepackage{array,multirow}
\usepackage{upgreek}
\usepackage{textcomp}
\usepackage{hyperref}
\usepackage[referable]{threeparttablex}
\usepackage{booktabs, dcolumn}

\newcommand{\beq}{\begin{equation}}
\newcommand{\eeq}{\end{equation}}
\newcommand{\simlt}{\mathrel{\hbox{\rlap{\hbox{\lower4pt\hbox{$\sim$}}}\hbox{$<$}}}}
\newcommand{\simgt}{\mathrel{\hbox{\rlap{\hbox{\lower4pt\hbox{$\sim$}}}\hbox{$>$}}}}

\newcommand{\Msol}{\;\mathrm{M}_{\odot}}

\newcommand{\Mbh}{\left(\frac{M_{\rm BH}}{10^{6}}\right)}

\newcommand{\rtidal}{\;r_{\rm t}}
\newcommand{\rg}{r_{\rm g}}
\newcommand{\physrad}{\mathcal{R}_{\rm t}}
\newcommand{\Ltsq}{\mathcal{L}_{\rm t}^2}

\def\apjl{ApJL}
\def\apj{ApJ}
\def\mnras{M.N.R.A.S.}
\def\aap{A\&A}
\def\nat{Nat.}

\def\apjs{ApJ Supp.}

\def\physrep{Phys. Rep.}

\def\prd{prd}

\defcitealias{Ryu1+2019}{Paper 1}
\defcitealias{Ryu2+2019}{Paper 2}
\defcitealias{Ryu3+2019}{Paper 3}

\newcommand{\harm}{{\sc Harm3d}}   
\newcommand{\mesa}{{\small MESA}}

\begin{document}

\title{Tidal disruptions of main sequence stars - IV. Relativistic effects and dependence on black hole mass}

\correspondingauthor{Taeho Ryu}
\email{tryu2@jhu.edu}

\author[0000-0002-0786-7307]{Taeho Ryu}
\affil{Physics and Astronomy Department, Johns Hopkins University, Baltimore, MD 21218, USA}

\author{Julian Krolik}
\affiliation{Physics and Astronomy Department, Johns Hopkins University, Baltimore, MD 21218, USA}
\author{Tsvi Piran}
\affiliation{Racah Institute of Physics, Hebrew University, Jerusalem 91904, Israel}

\author{Scott C. Noble}
\affiliation{Gravitational Astrophysics Laboratory, Goddard Space Flight Center, Greenbelt, MD 20771, USA}

\begin{abstract}
Using a suite of fully relativistic hydrodynamic simulations applied to main-sequence stars with realistic internal density profiles, we examine full and partial tidal disruptions across a wide range of black hole mass ($10^{5}\leq M_{\rm BH}/\mathrm{M}_{\odot}\leq 5\times 10^{7}$) and stellar mass ($0.3 \leq M_{\star} /\mathrm{M}_{\odot}\leq 3$) as larger $M_{\rm BH}$ leads to stronger relativistic effects.
For fixed $M_{\star}$, as $M_{\rm BH}$ increases, the ratio of the maximum pericenter distance yielding full disruptions ($\mathcal{R}_{\rm t}$) to its Newtonian prediction rises rapidly, becoming triple the Newtonian value for $M_{\rm BH} = 5\times10^{7}~{\rm M}_\odot$, while the ratio of the energy width of the stellar debris for full disruptions to the Newtonian prediction decreases steeply, resulting in a factor of two correction at $M_{\rm BH} = 5 \times 10^7~{\rm M}_\odot$. We provide approximate formulae that express the relativistic corrections of both $\mathcal{R}_{\rm t}$ and the energy wdith relative to their Newtonian approximate estimates. For partial disruptions, we find that  the fractional remnant mass for a given ratio of the pericenter to $\mathcal{R}_{\rm t}$ is higher for larger $M_{\rm BH}$. 

These results have several implications. As $M_{\rm BH}$ increases above $\sim 10^7~{\rm M}_\odot$, the  cross section for complete disruptions is suppressed by competition with direct capture.  However, the cross section ratio for partial to complete disruptions depends only weakly on $M_{\rm BH}$. 
The relativistic correction to the debris energy width delays the time of peak mass-return rate and diminishes the magnitude of the peak return rate.  For $M_{\rm BH} \gtrsim 10^7~{\rm M}_\odot$, the $M_{\rm BH}$-dependence of the full disruption cross section and the peak mass-return rate and time is influenced more by relativistic effects than by Newtonian dynamics.

\end{abstract}

\keywords{black hole physics $-$ gravitation $-$ hydrodynamics $-$ galaxies:nuclei $-$ stars: stellar dynamics}

\section{Introduction} \label{sec:intro}

Supermassive black holes (SMBHs) tidally disrupt stars when their separation becomes smaller than the so-called ``tidal radius''. Roughly half the mass removed from the star is bound to the black hole and may produce a luminous flare when it returns to the black hole, while the other half is expelled. 

Tidal disruption events (TDEs) caused by a $10^{6}\Msol$ SMBH have been considered a representative case in many theoretical studies \citep[e.g.,][]{Ayal+2000, Guillochon+2013, Mainetti+2017,Goicovic+2019}. However, in reality, TDEs can occur for a wide range of mass $M_{\rm BH}$.
It is therefore useful to study how the key properties of tidal disruptions depend on $M_{\rm BH}$.   The interest of this study is enhanced by the fact that Newtonian order of magnitude estimates suggest that the characteristic tidal radius measured in gravitational units, i.e., $r_{\rm t}/\rg \equiv (R_{\star}/r_{\rm g}) (M_{\rm BH}/M_{\star})^{1/3} \propto M_{\rm BH}^{-2/3}$, where $R_{\star}$ is the stellar radius, $M_{\star}$ is the stellar mass and $r_{\rm g}$ is the gravitational radius, $r_{\rm g}=G M_{\rm BH}/c^{2}$.
Given that scaling, these events take place in increasingly relativistic environments as $M_{\rm BH}$ increases.  A study of black hole mass-dependence is therefore a study of how relativistic effects alter the course of these events (see a recent review by \citealt{Stone+2019} for TDEs in relativity).

We aim to accomplish this study by performing relativistic hydrodynamic simulations (using \harm: \citealt{Noble+2009}) whose initial conditions are realistic main-sequence stellar models taken from the stellar evolution code \mesa.
In particular, we will examine a small sample of stellar masses ($0.3 \Msol$, $1.0\Msol$, and $3.0 \Msol)$ being disrupted by black holes of six different masses: $10^{5}\Msol$, $10^{6}\Msol$, $5\times10^{6}\Msol$, $10^{7}\Msol$, $3\times10^{7}\Msol$ and $5\times10^{7}\Msol$. In Section~\ref{sec:results}, we present results for the physical tidal radius $\physrad$ (Section~\ref{sec:physicalR_BH}), the energy distribution of stellar debris and the resulting fallback rate (Section~\ref{sub:energydistribution}), and the remnant mass of partial disruptions (Section~\ref{sub:remnantmass}).
In Section~\ref{sec:implication}, we discuss the TDE event rate (Section~\ref{sub:TDErate}). We also reconsider the maximum black hole mass for tidal disruptions (Section~\ref{sub:max_mBH}).
Lastly, we summarize our findings in Section~\ref{sec:summary}.

Throughout the remainder of this paper, all masses will be measured in units of ${\rm M}_\odot$ and all stellar radii in units of ${\rm R}_\odot$.

\begin{table}
 	\renewcommand{\thetable}{\arabic{table}}
 	\centering
 	\caption{Values of $r_{\rm p}/r_{\rm t}$ considered in these experiments. The units of $M_{\star}$ and $M_{\rm BH}$ are $\rm{M}_{\odot}$. We also show the range of the ``penetration factor" $\beta$.} \label{tab:psi}
 	\begin{tabular}{c c c c}
 		\tablewidth{0pt}
 		\hline
 		\hline
 		$M_{\star}$&  $M_{\rm BH}[10^{6}]$   & 	$r_{\rm p}/r_{\rm t}$ & $\beta \equiv \rtidal/r_{\rm p}$\\
 		\hline
 		\multirow{4}{*}{ $0.3$}          &  $0.1$  & 1.0, 1.1, 1.2, 1.3, 1.5, 1.7 & [0.59,~1.0] \\  
 		                                 &  $1$  & 1.0, 1.2, 1.3, 1.4, 1.5, 1.8  & [0.56,~1.0]\\  
 		                                 &  $5$  & 1.2, 1.6, 1.7, 1.8, 2.0, 2.1 & [0.48,~0.83]\\  
 		                                 &  $10$   & 1.8, 1.9, 2.0, 2.1, 2.1, 2.3& [0.43,~0.56]\\  
 		                                 &  $30$   & 2.5, 2.6, 2.7, 2.75, 2.8, 3.0& [0.33,~0.40]\\  
 		                                 &  $50$   & 3.0, 3.1, 3.2, 3.3, 3.4, 3.5& [0.29,~0.33] \\   		                                 \hline
 		\multirow{4}{*}{ $1.0$}          &  $0.1$  & 0.40, 0.45, 0.50, 0.55, 0.65, 0.80& [1.3,~2.5]\\  
 		                                 &  $1$  & 0.40, 0.45, 0.50, 0.55, 0.65, 1.00& [1.0,~2.5]\\  
 		                                 &  $5$  & 0.5, 0.6, 0.7, 0.8, 0.9, 1.4& [0.71,~1.0]\\  
 		                                 &   $10$  & 0.6, 0.8, 0.9, 1.0, 1.1, 1.3& [0.77,~1.7]\\ 
 		                                 &  $30$   & 1.0, 1.1, 1.2, 1.3, 1.5, 1.6& [0.63,~1.0]\\  
 		                                 &  $50$   & 1.2, 1.4, 1.5, 1.55, 1.6, 1.7& [0.59,~0.83]\\   		            \hline 
 		\multirow{4}{*}{ $3.0$}          &  $0.1$  & 0.35, 0.40, 0.45, 0.60, 0.8, 1.0& [1.0,~2.9]\\  
 		                                 &  $1$  & 0.35, 0.40, 0.45, 0.50, 0.60, 0.85& [1.2,~2.9]\\  
 		                                 &  $5$  & 0.4, 0.5, 0.6, 0.7, 0.8, 1.0& [1.0,~2.5]\\  
 		                                 &   $10$   & 0.5, 0.6, 0.7, 0.8, 1.0, 1.2& [0.83,~2.0]\\  
 		                                 &  $30$   & 0.7, 0.8, 0.9, 1.0, 1.1, 1.3& [0.77,~1.4]\\  
 		                                 &  $50$   & 0.8, 0.95, 1.05, 1.1, 1.2, 1.3& [0.77,~1.3]\\   		            		                                 \hline
 		\hline
 	\end{tabular}
 \end{table}

 \begin{table*}
 	\renewcommand{\thetable}{\arabic{table}}
 	\centering
 	\caption{The physical tidal radii $\physrad$ for different $M_{\rm BH}$, in units of $r_{\rm g}$; the specific angular momentum $\mathcal{L}_{\rm t}\equiv L(\physrad)$, in units of $r_{\rm g}c$; $\physrad/r_{\rm t}(\equiv \Psi)$ and $\beta_{\rm d}(\equiv \Psi^{-1})$. The units of $M_{\star}$ and $M_{\rm BH}$ are $M_{\star}$.} \label{tab:tidal_r_BH}
 	\begin{tabular}{c| c c c c c c}
 		\tablewidth{0pt}
 		\hline
 		\hline
 		$M_{\rm BH}$  &  $10^{5}$  &  $10^{6}$ & $5\times10^{6}$ & $10^{7}$  & $3\times10^{7}$ & $5\times10^{7}$\\\hline
 		&  \multicolumn{6}{c}{$\physrad/r_{\rm g}$}   \\
 		 $M_{\star}=0.3$&   $113\pm5$ & $26.5\pm1.1$& $12.0\pm0.4$ & $8.9\pm0.2$ & $5.8\pm0.1$ & $4.9\pm0.1$\\
	    $M_{\star}=1.0$  &   $93.6\pm5.5$ & $22.5\pm1.2$& $10.5\pm0.8$ & $8.7\pm0.5$ & $5.7\pm0.2$ & $5.1\pm0.2$	\\
     	$M_{\star}=3.0$  &$139\pm9$ & $33.9\pm2.0$ & $15.0\pm1.4$ & $11.2\pm0.9$ & $7.0\pm0.4$ & $5.9\pm0.3$\\
 		\hline\hline
            & 	\multicolumn{6}{c}{$\mathcal{L}_{\rm t}/(r_{\rm g}c)$}\\
    $M_{\star}=0.3$         &	$15.2\pm0.3$&  $7.58\pm0.14$  &  $5.36\pm0.06$  & $4.80\pm0.04$ & $4.21\pm0.02$ & $4.07\pm0.01$ \\
  $M_{\star}=1.0$  &	$13.8\pm0.4$&  $7.03\pm0.17$  &  $5.10\pm0.15$ &$4.75\pm0.10$ &$4.18\pm0.04$ & $4.10\pm0.02$ \\      
  	$M_{\star}=3.0$ &	$15.8\pm0.6$&  $8.49\pm0.23$  &  $5.89\pm0.23$ & $5.22\pm0.16$ &$4.43\pm0.08$ & $4.22\pm0.05$\\
  	     	\hline\hline
            & 	\multicolumn{6}{c}{$\physrad/r_{\rm t}(\equiv\Psi)$}\\
    $M_{\star}=0.3$         &	$1.15\pm0.05$&  $1.25\pm0.05$  &  $1.65\pm0.05$  & $1.95\pm0.05$ & $2.65\pm0.05$ & $3.15\pm0.05$ \\
  $M_{\star}=1.0$  &	$0.425\pm0.025$&  $0.475\pm0.025$  &  $0.65\pm0.05$ &$0.85\pm0.05$ &$1.15\pm0.05$ & $1.45\pm0.05$ \\      
  	$M_{\star}=3.0$ &	$0.375\pm0.025$&  $0.425\pm0.025$  &  $0.55\pm0.05$ & $0.65\pm0.05$ &$0.85\pm0.05$ & $1.00\pm0.05$\\
  	     	\hline\hline
            & 	\multicolumn{6}{c}{$\beta_{\rm d}(\equiv\Psi^{-1})$}\\
    $M_{\star}=0.3$         &	$0.87\pm0.04$&  $0.80\pm0.03$  &  $0.61\pm0.02$  & $0.51\pm0.01$ & $0.38\pm0.01$ & $0.32\pm0.01$ \\
  $M_{\star}=1.0$  &	$2.35\pm0.14$&  $2.11\pm0.11$  &  $1.54\pm0.12$ &$1.17\pm0.07$ &$0.87\pm0.04$ & $0.69\pm0.02$ \\      
  	$M_{\star}=3.0$ &	$2.67\pm0.18$&  $2.35\pm0.14$  &  $1.82\pm0.17$ & $1.54\pm0.12$ &$1.18\pm0.07$ & $1.00\pm0.05$\\
  	
\hline
 	\end{tabular}
 \end{table*}

 \section{Simulations}
 \label{sec:sim_setup}

Our simulations differ from those described in \citetalias{Ryu2+2019} and \citetalias{Ryu3+2019} only by using a wider range of black hole masses: $10^{5}\Msol$, $10^{6}\Msol$, $5\times10^{6}\Msol$, $10^{7}\Msol$, $3\times10^{7}\Msol$ and $5\times10^{7}\Msol$.  In all cases we use the fully general relativistic hydrodynamics code \harm~\citep{Noble+2009} operating in a Schwarzschild spacetime, but in a coordinate frame we call the box frame that follows the star's center-of-mass trajectory.

The initial internal structure of each star is taken from a \mesa~model at an age equal to half its main-sequence lifetime \citep{Paxton+2011}.  The case with mass $M_{\star}=0.3$ represents fully convective stars; $M_{\star}=1$ is our example of a (nearly) fully radiative star; like other high-mass stars, $M_{\star}=3$ is radiative outside a convective core (see their density profiles in \citetalias{Ryu2+2019}).  The choice of these three masses was motivated by the fact that for $M_{\rm BH}=10^{6}$, $\physrad$ for $0.15\leq M_{\star}\leq3$ is bounded below by its value for $1\Msol$ and bounded above by its value for $3\Msol$, while $\physrad$ for $0.3\Msol$ is closest to the average value ($\physrad\simeq27~\rg$) within the range of masses $0.15\leq M_{\star}\leq3$ (\citetalias{Ryu1+2019}). As we showed in \citetalias{Ryu1+2019}, relativistic corrections to ${\cal R}_{\rm t}$ are almost independent of $M_{\star}$.  This fact suggests that these three masses should play the same roles (average, lower, and upper bound) for any $M_{\rm BH}$.

Although the background spacetime is fully relativistic,  the star's self-gravity is calculated using a Newtonian Poisson solver in a frame comoving with the star defined by a tetrad system at the star's center-of-mass.  In this frame, the metric is exactly Minkowski at the origin, but deviates from Minkowski elsewhere (see \citetalias{Ryu2+2019} for details). The approximation of Newtonian self-gravity is valid  when  both the  self-gravity and, more importantly, the non-Minkowski terms associated with tidal gravity, are small throughout the simulation volume.  This criterion is satisfied in the tetrad frame, but not in the box frame.  The stellar potential is added to $g_{\rm tt}$ in the tetrad frame as a well-justified post-Newtonian approximation because in relativistic units it is $\lesssim 10^{-6}$.  To obtain the metric in the box frame, we then apply an inverse tetrad transformation. Quantitative limits for the applicability of this approximation are presented in Appendix~A in \citetalias{Ryu2+2019}.  As remarked in \citet{Ryu2+2019}, if stellar self-gravity is added to $g_{\rm tt}$ in the box frame, where tidal gravity is significant, rather than in the tetrad frame, errors in the gravitational acceleration at the tens of percent level can be created.
Although the departure of the background metric from Minkowski grows as the separation to the BH falls, these departures are always small in our simulations.  Even along the outer edges of the simulation box, where they are largest, at a distance from the black hole $\simeq 100r_{\rm g}$ they are $\sim 10^{-4}$ and rise to only $\sim 10^{-2}$ at $\simeq 5r_{\rm g}$.

For each stellar mass, we performed a suite of simulations for TDEs with various pericenter distances $r_{\rm p}/r_{\rm t}$ separated by increments $r_{\rm p}/r_{\rm t}=0.05-0.25$. We tabulate the values of $r_{\rm p}/r_{\rm t}$  considered in these experiments in Table~\ref{tab:psi}. The quantity $r_{\rm p}/r_{\rm t}$ is the inverse of the ``penetration factor" $\beta$.

To distinguish full from partial disruptions, we employ the same criteria introduced in \citetalias{Ryu2+2019}, i.e., requiring full disruptions to have:
\begin{enumerate}
\item \label{con1} No approximately-spherical bound structure. 
\item \label{con2}  Monotonic (as a function of time) decrease in the maximum pressure of the stellar debris. 
\item \label{con3} Monotonic (as a function of time) decrease in the mass within the computational box. 
\end{enumerate} 	
We refer to events satisfying all of those conditions as ``full'', others we call ``partial". We estimate the physical tidal radius $\physrad$, the maximal radius at which a full tidal disruption takes place, as the mean of the greatest $r_{\rm p}$ yielding a full disruption and the smallest $r_{\rm p}$ producing a partial disruption. The uncertainty in $\mathcal{R}_{\rm t}$ is due to our discrete sampling of $r_{\rm p}$.

\begin{figure}
	\centering
	\includegraphics[width=8.9cm]{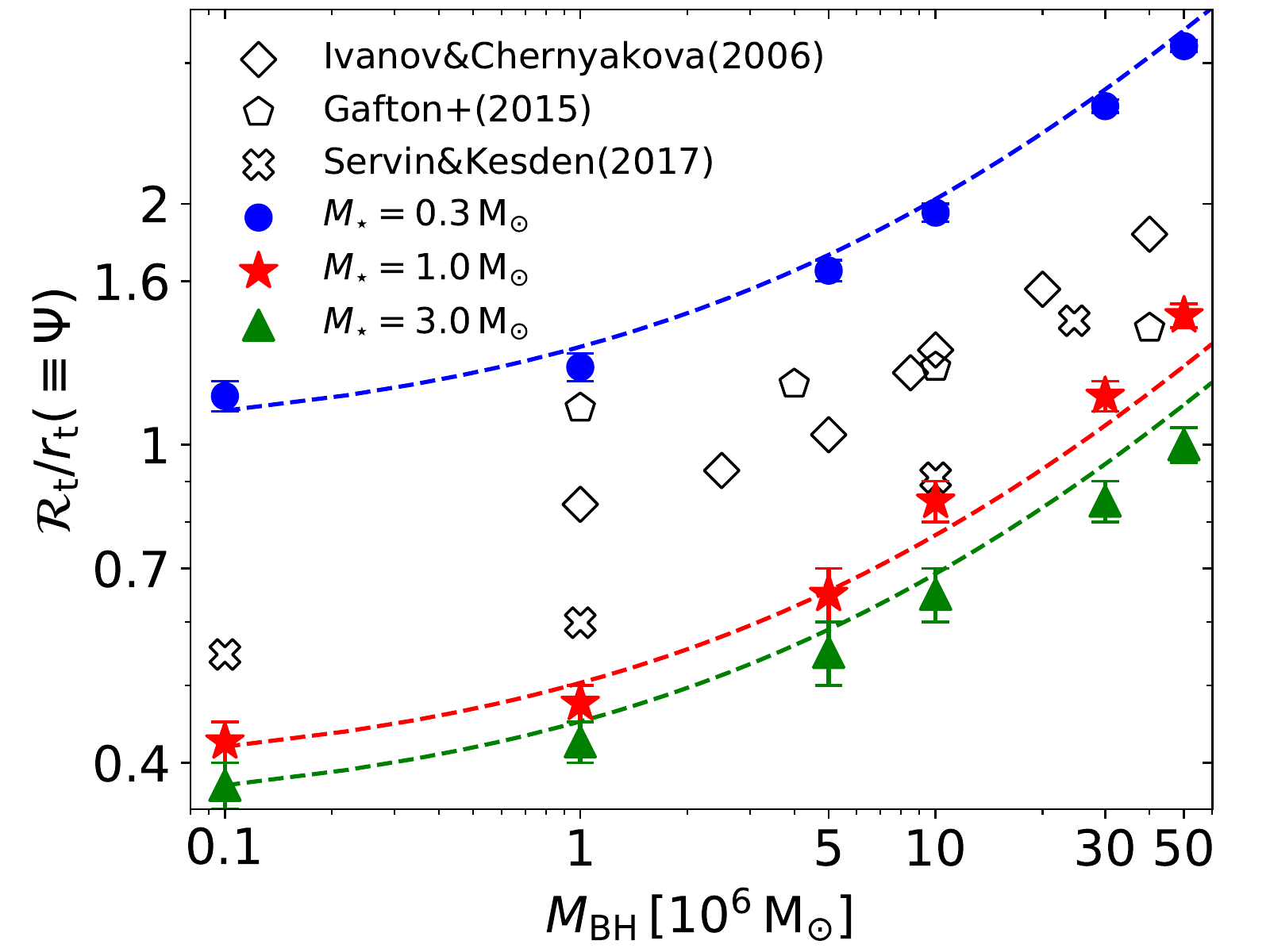}	
	\caption{The  physical tidal radius in units of the nominal tidal radius, $\physrad/r_{\rm t}(\equiv\Psi)$, shown by filled symbols color-coded to indicate mass as shown in the legend. The curves indicate the fitting formula (Equation~\ref{eq:fit_psibh}), multiplied by $\Psi(M_{\star},M_{\rm BH}=10^6)$. The hollow symbols show $\Psi$ for $M_{\star}=1$ from \citet{IvanovChernyakova2006} (diamonds), \citet{Gafton+2015} (pentagons) and \citet{ServinKesden2017} (crosses).}
	\label{fig:r_tidal_psi}
\end{figure}

 \section{Results}
\label{sec:results}

\subsection{Physical tidal radius $\physrad$}
\label{sec:physicalR_BH}

The physical tidal radius $\mathcal{R}_{\rm t}$ is the maximum radius within which a full tidal disruption takes place.  The actual values measured in our numerical experiments are tabulated in Table~\ref{tab:tidal_r_BH}.
Figure~\ref{fig:r_tidal_psi} illustrates them graphically, showing $\physrad/r_{\rm t}(\equiv\Psi)$ as a function of $M_{\rm BH}$ for the three stellar models.  For comparison, it also shows the equivalent predictions of two other studies employing relativistic calculations of the tidal stresses.  As can be seen easily, both $\Psi$ and $d\Psi/dM_{\rm BH}$ increase with greater $M_{\rm BH}$.  Tidal forces are more destructive as relativistic effects become more significant, which leads to larger $\Psi$.  From the Newtonian limit ($M_{\rm BH}=10^{5}$) to the strongly relativistic conditions of $M_{\rm BH} = 5 \times 10^7$, $\Psi$ grows by a factor $\sim 3$.

Figure~\ref{fig:r_tidal_psi} also shows the $M_{\rm BH}$-dependence of $\Psi$ has only a weak dependence on $M_{\star}$ (also see the \textit{left} panel of Figure 2 in \citetalias{Ryu1+2019}). This fact allows us to find an analytic expression for the $M_{\rm BH}$-dependence of $\Psi$ separate from that for the $M_{\star}$-dependence. The expression for the $M_{\rm BH}$-dependent term, which we call $\Psi_{\rm BH}$ in \citetalias{Ryu1+2019}, is, 
\begin{equation}
\Psi_{\rm BH}(M_{\rm BH})=0.80 + 0.26~\left(\frac{M_{\rm BH}}{10^{6}}\right)^{0.5},\label{eq:fit_psibh}
\end{equation}
which is depicted in Figure~\ref{fig:r_tidal_psi} using dashed lines.
By comparing the logarithmic derivative of $\Psi_{\rm BH}$ with respect to $M_{\rm BH}$ (i.e. $d\ln \Psi_{\rm BH}/d\ln M_{\rm BH}>1$), we find that for black holes more massive than $\sim 3 \times 10^7 \Msol$, the size of the physical tidal radius is more sensitive to relativistic corrections than to the simple Newtonian comparison of stellar self-gravity to black hole tidal gravity.

Several previous efforts have also explored this trend, \citet{IvanovChernyakova2006}, \citet{Gafton+2015} and \citet{ServinKesden2017}, which are indicated using hollow symbols in Figure~\ref{fig:r_tidal_psi}.  All sought to explore relativistic effects in TDEs, but did so with a variety of approximations. \citet{IvanovChernyakova2006} calculated the tidal stress exactly, but described their star as a set of ellipsoidal shells whose initial structure was that of a $M_{\star}=1$ $\gamma=5/3$ polytrope (i.e., having the internal density profile of a low-mass star), and whose pressure and self-gravity were computed in a 1$-d$ approximation.  
\citet{Gafton+2015} employed a ``generalized Newtonian potential''  \citep{TejedaRosswog2013} that reproduces test-particle motion in a Schwarzschild spacetime very well when the specific energy is unity; it is unclear how well it reproduces relativistic tidal stresses and debris motion.  Their stars were supposed to be $\gamma=5/3$ polytropes with $1\Msol$, and the stellar self-gravity was computed in an entirely Newtonian fashion. \citet{ServinKesden2017} constructed an analytic expression for mapping Newtonian hydrodynamics simulations of $\gamma = 4/3$ polytropes with $M_{\star}=1$ to Schwarzschild geodesics by matching the magnitude of the tidal stresses at pericenter.  As shown in Figure~\ref{fig:r_tidal_psi}\footnote{The data plotted were read from Figure~5 in \citealt{IvanovChernyakova2006}, Figure~3 in \citealt{Gafton+2015} and Figure~8 in \citealt{ServinKesden2017}.}, the alteration to the tidal radius due solely to relativistic effects found by the first and third efforts \citep{IvanovChernyakova2006, ServinKesden2017} is similar to ours, but \citet{Gafton+2015} found a weaker dependence on $M_{\rm BH}$.  Because relativity enters this part of the problem largely through the tidal stress, this should, perhaps, be unsurprising.

Where the results of \citet{IvanovChernyakova2006} and \citet{ServinKesden2017} differ from ours, as well as each other's, is in the normalization.  Compared to our results for $M_{\star}=1$, $\Psi$ from \citet{IvanovChernyakova2006} is $50-80\%$ larger, while the predictions of \citet{ServinKesden2017} are closer to ours, $10-30\%$ larger.  The closer agreement with \citet{ServinKesden2017} is likely due to the coincidence that $\gamma=4/3$, although physically inappropriate, produces a good approximation to the density profile of a realistic main sequence star with $M_{\star}=1$.

  Lastly we note that  \cite{Tejeda+2017} and \cite{Gafton2019} used a relativistic hydrodynamics SPH code with Newtonian self-gravity to probe the relativistic regime. Their study employed a $\gamma=5/3$ polytrope for $M_{\star}=1$ stars and considered how the encounters depended on $\beta$ and spin parameter $a/M$ for  a single black hole mass, $M_{\rm BH}= 10^6$, paying special attention to debris geometry due to black hole spin. In contrast, we have determined how the tidal disruption properties of realistic main sequence stars depend on $M_{\rm BH}$ over a wide range of masses.

\begin{figure}
	\centering
	\includegraphics[width=8.8cm]{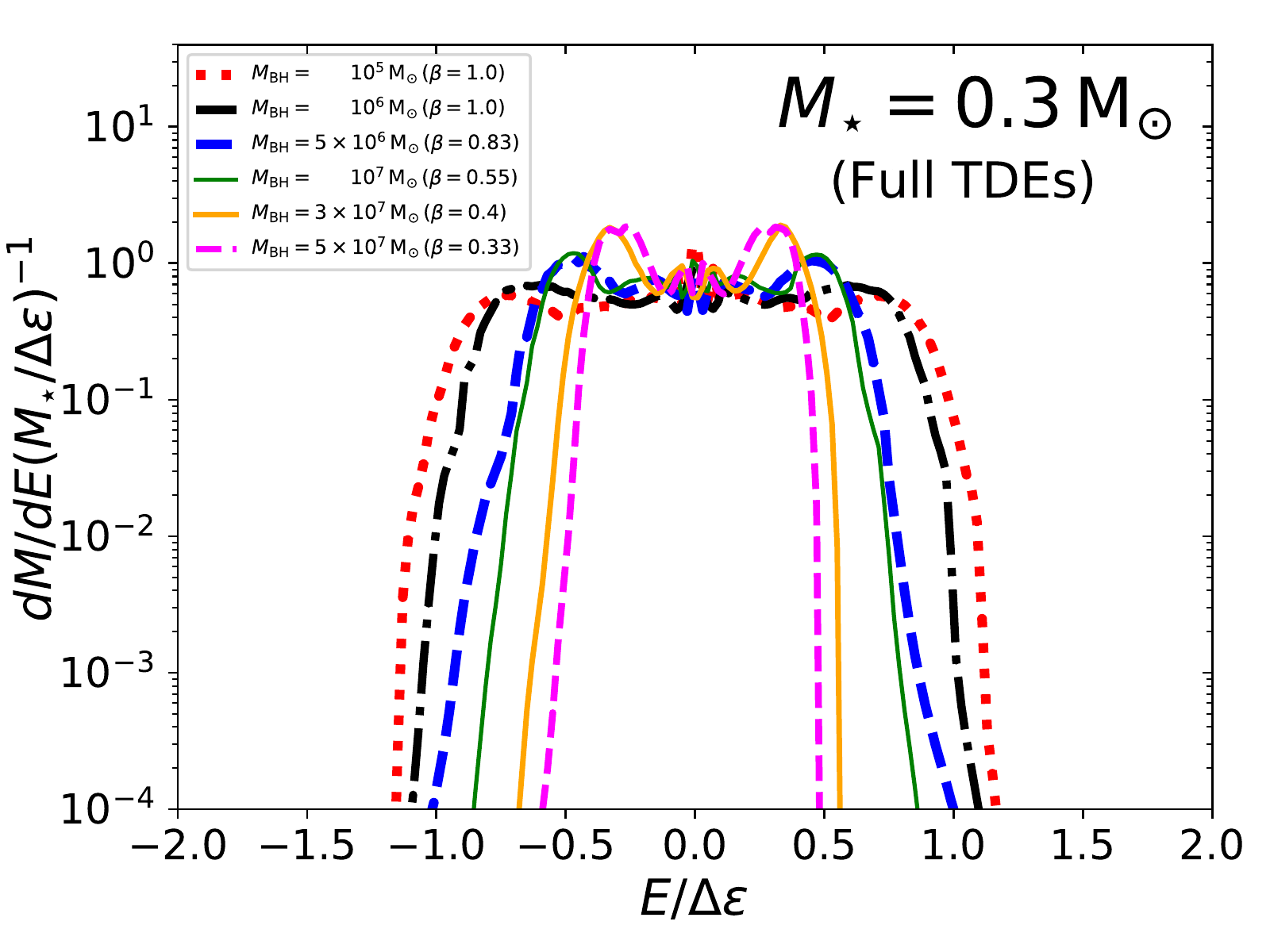}
	\includegraphics[width=8.8cm]{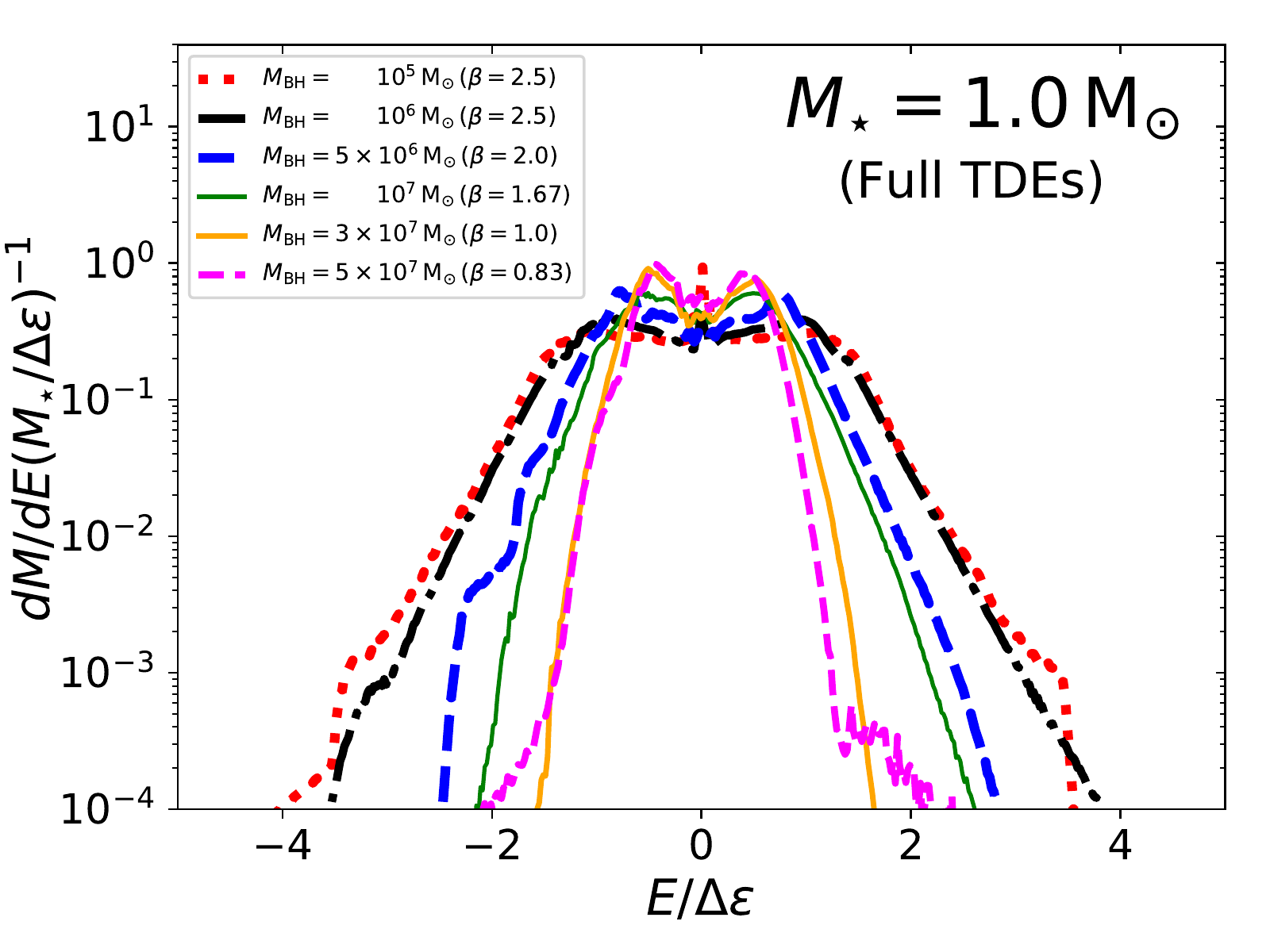}	
	\includegraphics[width=8.8cm]{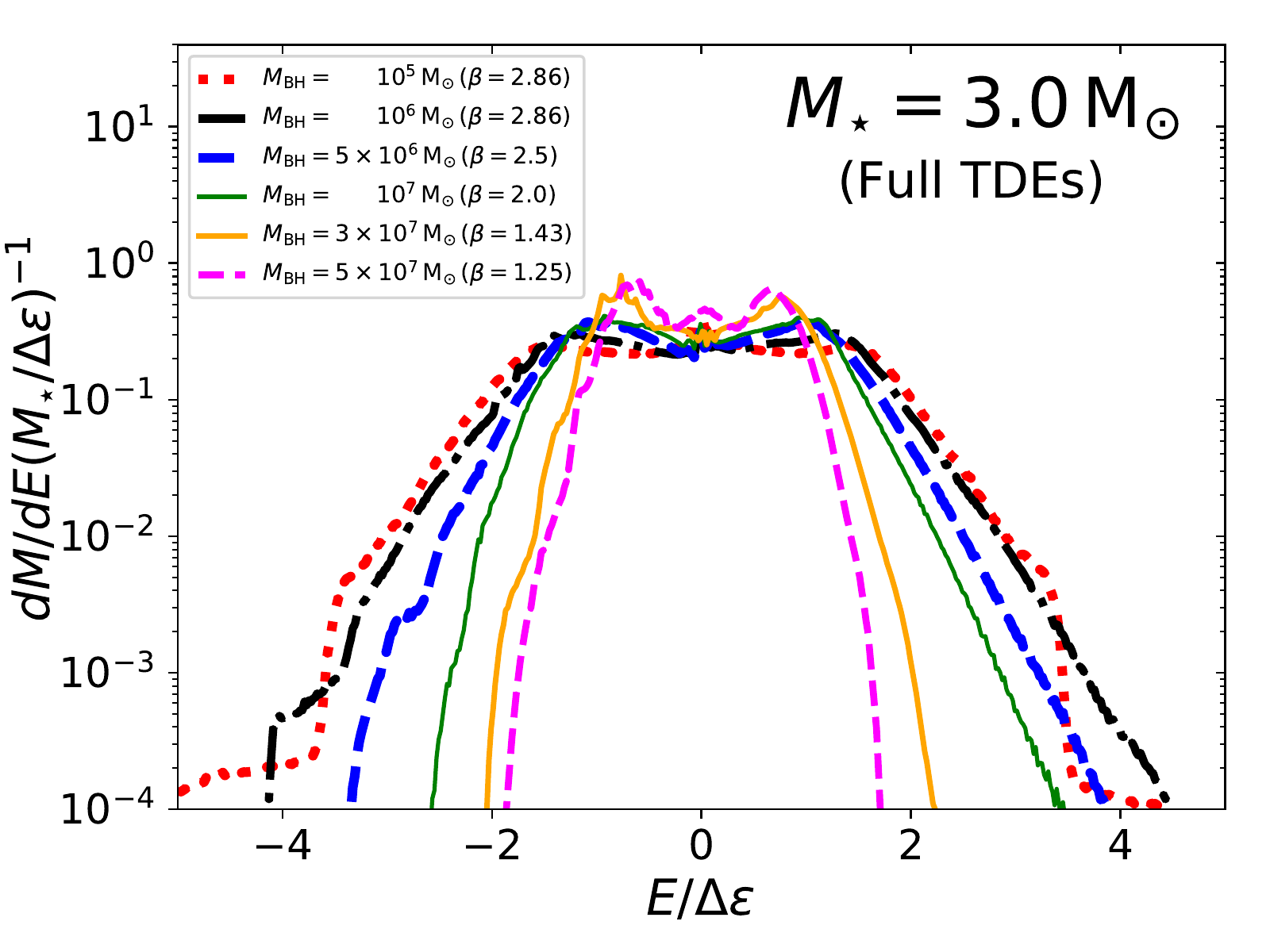}		
	\caption{The energy distribution $dM/dE$ for full disruptions of $M_{\star}=0.3$ (\textit{top} panel), $1.0$ (\textit{middle} panel) and $3.0$  (\textit{bottom} panel). }
	\label{fig:dm_de}
\end{figure}

\subsection{Energy distribution and fallback rate of stellar debris for full disruptions}
\label{sub:energydistribution}

The energy distribution of stellar debris directly determines their orbits. In the conventional description of TDEs \citep{Rees1988}, the energy distribution $dM/dE$ is  approximated as  flat within a characteristic energy width $\pm\Delta E$. In relativistic language, the classical specific orbital energy $E \equiv -u_{\rm t} - 1$ evaluated in the black hole frame, i.e., it is the conserved relativistic specific orbital energy exclusive of the rest mass energy. This characteristic width is often estimated \citep{Lacy+1982,Stone+2013} as
\begin{align}
\Delta\epsilon= 
\frac{GM_{\rm BH}R_{\star}}{r_{\rm t}^{2}}.
\label{eq:deltae}
\end{align}
In this section, we focus on how $dM/dE$ varies as a function of $M_{\rm BH}$. 

Figure~\ref{fig:dm_de} shows $dM/dE$ for all 18 combinations of $M_{\star}$ and $M_{\rm BH}$. For all $M_{\star}$, $dM/dE$ becomes narrower and the ``shoulders'' (local maxima near the outer edges) become more conspicuous for higher $M_{\rm BH}$. As a result, the energy width $\Delta E$ containing 90\% of the total mass, when measured in units of $\Delta\epsilon$ is smaller for higher $M_{\rm BH}$ (see also Figure~5 in \citetalias{Ryu1+2019}), with small variations ($<5-10\%$) within the range of $r_{\rm p} < \physrad$ considered. In \citetalias{Ryu1+2019}, we provide an analytic expression for the $M_{\rm BH}$-dependence of $\Delta E/\Delta\epsilon(\equiv \Xi_{\rm BH})$, 
\begin{align}\label{eq:Xi1}
\Xi_{\rm BH}=1.27 - 0.300\Mbh^{0.242}.
\end{align}

In \citetalias{Ryu1+2019}, we also showed that $\Xi_{\rm BH}$ could be more crudely, but more simply, approximated by $\Psi_{\rm BH}^{-1}$.  It is interesting that any prediction for a spread in energy due solely to the tidal potential would have suggested this dependence would have been $\propto \Psi_{\rm BH}^{-2}$ rather than $\propto \Psi_{\rm BH}^{-1}$.   This is yet another piece of evidence supporting the argument given in \citetalias{Ryu1+2019} that the ``frozen-in" approximation is not a good basis on which to predict the debris energy spread.

Unlike $\Delta E$, the shape of the outer edge of the energy distribution depends on $M_{\rm BH}$ in a way that does depend on stellar mass. The distributions $dM/dE$ for $M_{\star}=1$ and $M_{\star}=3$ have significant tails for low $M_{\rm BH}$, but these become narrower for larger $M_{\rm BH}$. In contrast, $dM/dE$ for $M_{\star}=0.3$ has very sharp edges for the entire range of $M_{\rm BH}$.
Because Newtonian gravity is scale-free, it would not predict any changes in the shape of $dM/d(E/\Delta \epsilon)$ as a function of $M_{\rm BH}$;  only in general relativity, for which there is a special spatial scale and ${\cal R}_{\rm t}/r_{\rm g}$ is a function of $M_{\rm BH}$, can these trends emerge.

\begin{figure}
	\centering
	\includegraphics[width=8.8cm]{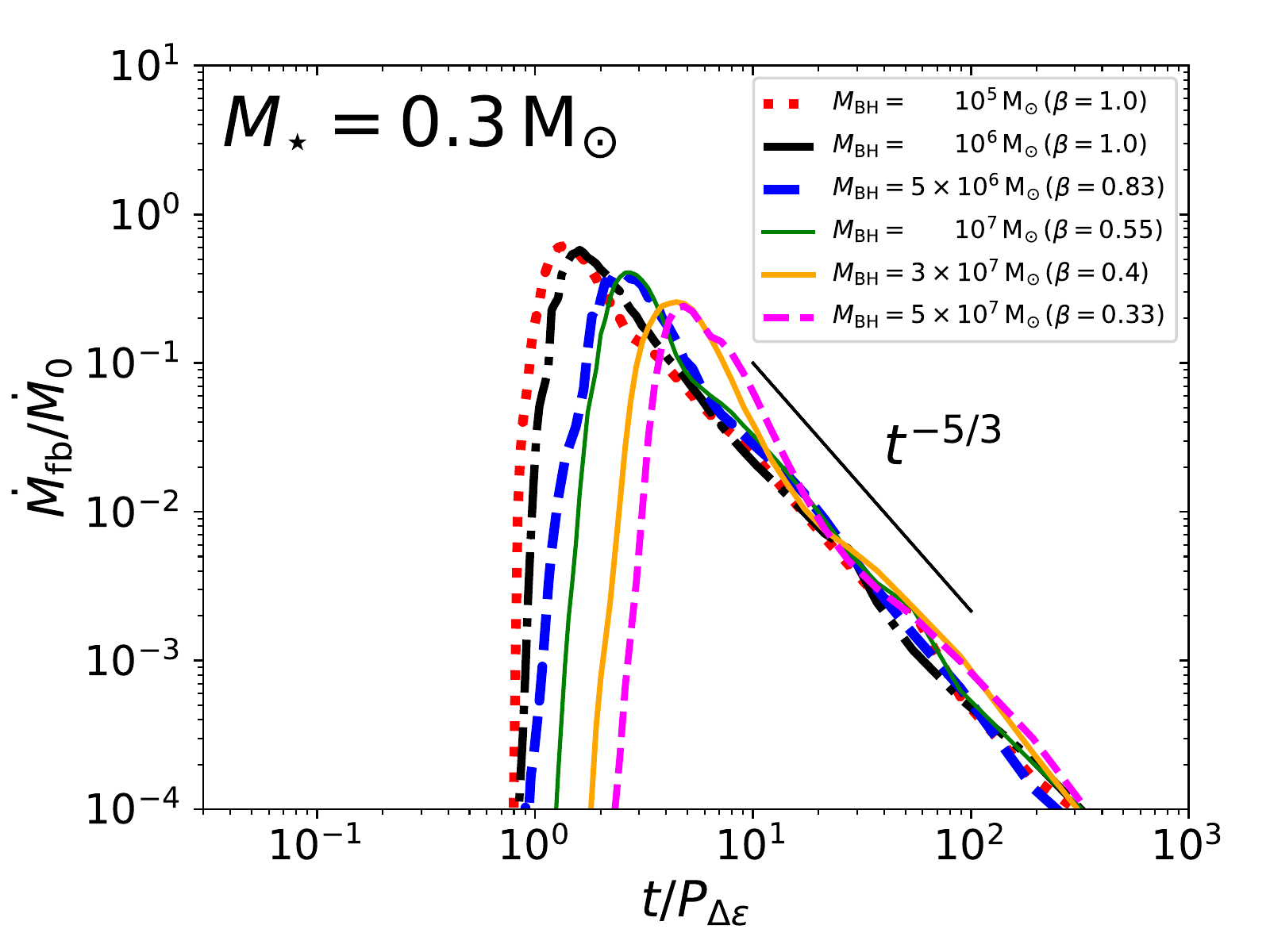}
	\includegraphics[width=8.8cm]{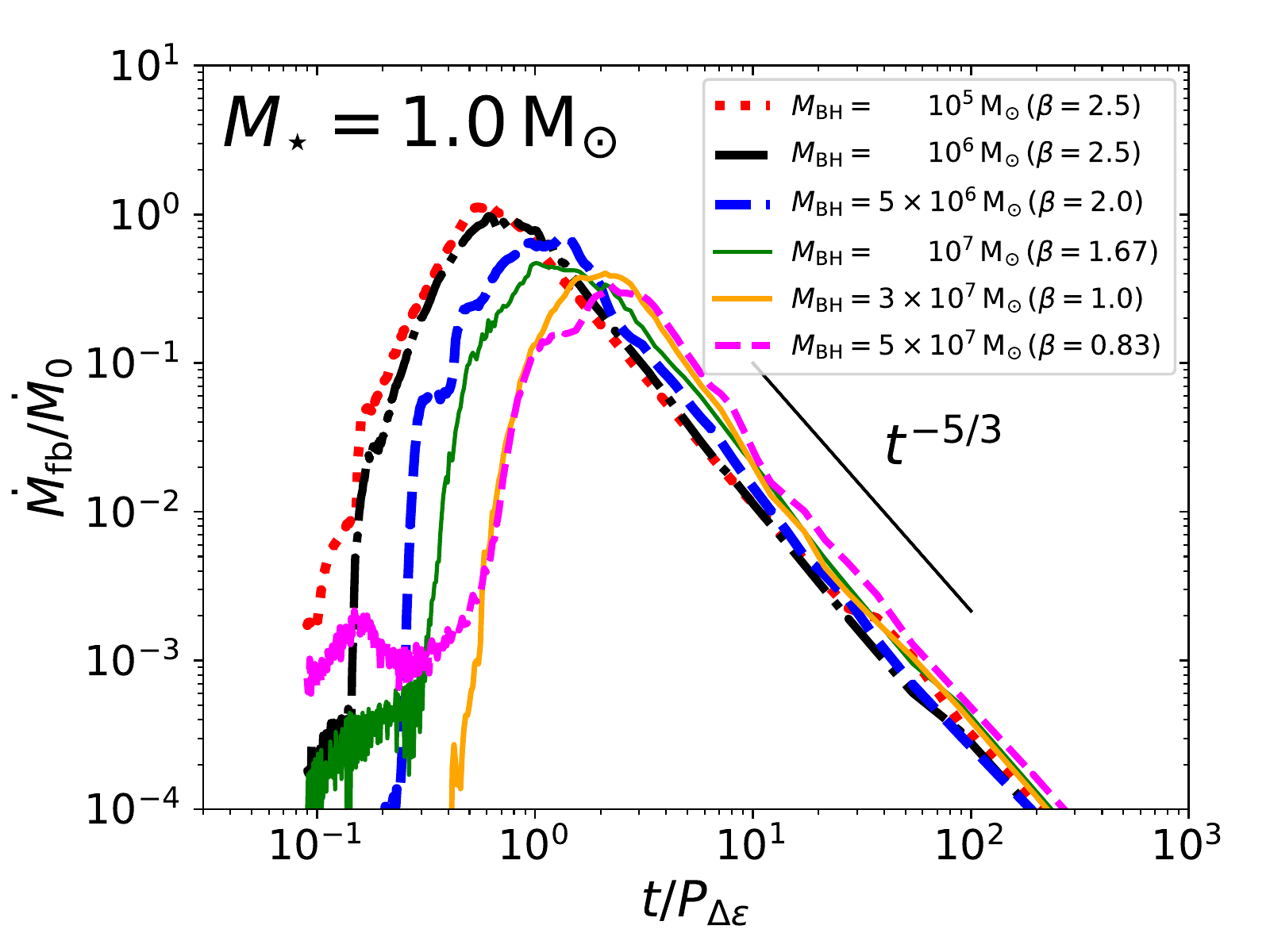}	
	\includegraphics[width=8.8cm]{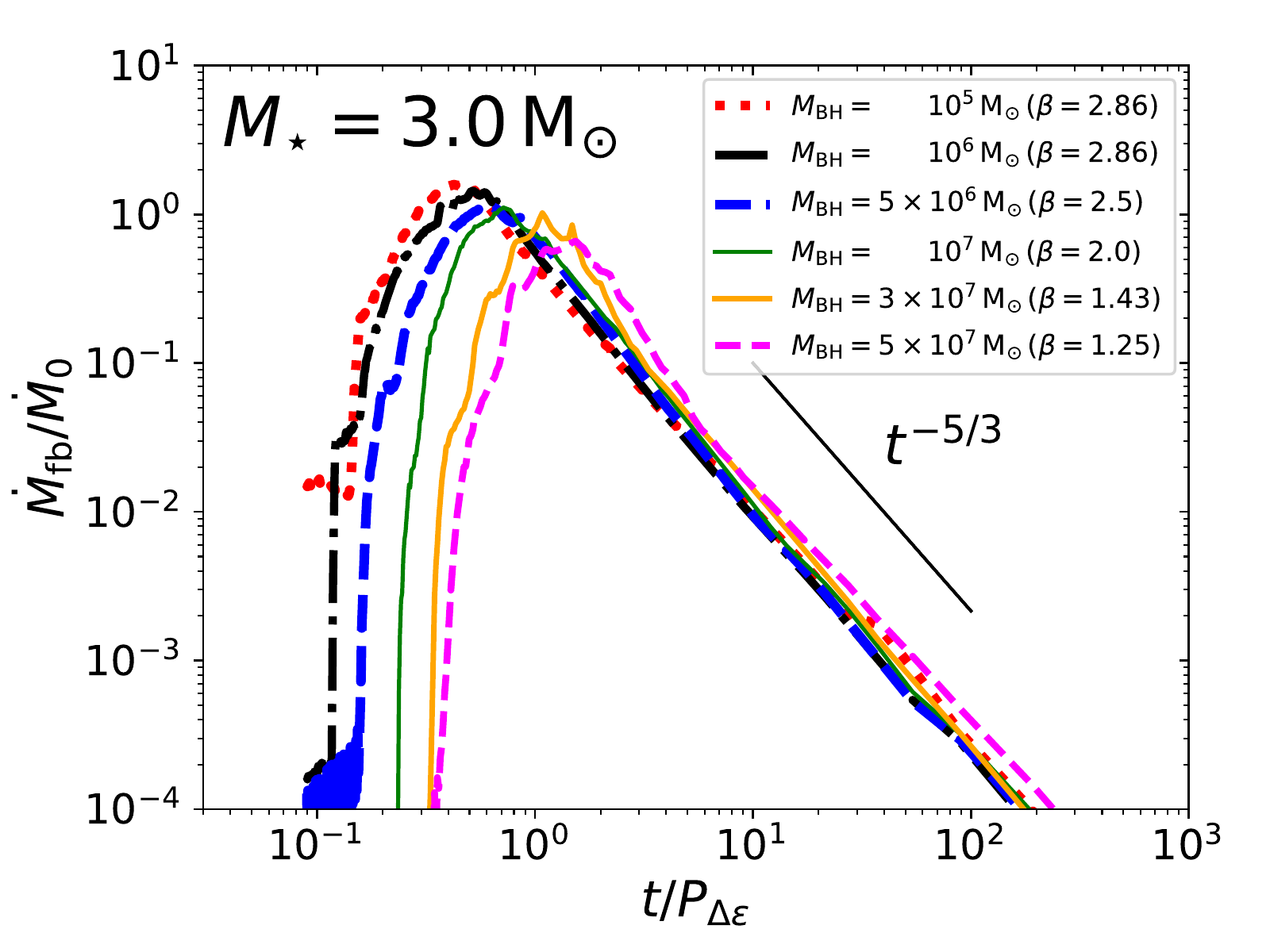}		
	\caption{The mass fall back rate $\dot{M}_{\rm fb}$ for $M_{\star}=0.3$ (\textit{top} panel), $1.0$ (\textit{middle} panel) and $3.0$ (\textit{bottom} panel), using the energy distributions shown in Figure~\ref{fig:dm_de}. The time and rate are normalized by $P_{\Delta\epsilon}$ and $\dot{M}_{0}=M_{\star}/3P_{\Delta\epsilon}$, respectively. Here, $P_{\Delta\epsilon}$ is the orbital period for the specific orbital energy of $\Delta\epsilon$. The diagonal line in each panel indicates the $t^{-5/3}$ power-law. }
	\label{fig:fallbackrate}
\end{figure}

\citet{ServinKesden2017} have also estimated the change in energy spread due to relativistic effects. Phrased in terms of our language, they assumed that the energy distribution is zero for $|E| > GM_{\rm BH}R_{\star}/\physrad^{2}$ and a constant value for $|E| \leq GM_{\rm BH}R_{\star}/\physrad^{2}$.   However, as we have seen, the character of the energy distribution is more complicated than a simple square wave, and its characteristic width is not $\propto \Psi_{\rm BH}^{-2}$ as this assumption would predict.  For reasons like these, and because mass-loss takes place across a wide span of radii at which stellar gravity, hydrodynamic forces, and tidal gravity are all competitive (\citetalias{Ryu2+2019}), approximating the energy spread in terms of the potential energy range at a particular location is not a particularly good approximation (\citetalias{Ryu1+2019}).

Using the expression for the mass fallback rate of stellar debris on ballistic orbits \citep{Rees1988,Phinney1989}, 
\begin{align}
\dot{M}_{\rm fb}&=\frac{dM}{dE}\left|\frac{dE}{dt}\right|=\frac{(2\uppi G M_{\rm BH})^{2/3}}{3}\frac{dM}{dE}t^{-5/3},
\label{eq:mdot}
\end{align}
and the energy distributions for the full disruptions in Figure~\ref{fig:dm_de}, we determine the mass fallback rate as a function of time. The results are depicted in Figure~\ref{fig:fallbackrate}, where the rate and time are normalized by $\dot{M}_{0} \equiv M_{\star}/(3P_{\Delta\epsilon})$ and $P_{\Delta\epsilon} \equiv \frac{\uppi}{\sqrt{2}}G M_{\rm BH}{\Delta\epsilon}^{-3/2}$, respectively. The shapes of the fallback curves are all qualitatively similar, possessing a rapid rise and a decline that is not far from the classical expectation, $\propto t^{-5/3}$.

However, it is also clear that, as a consequence of the decrease in $\Delta E$ with increasing $M_{\rm BH}$, the time at which the peak is reached increases for larger black holes and the associated fallback rate decreases (because for any given $M_{\star}$, the total amount of mass returning is fixed).  The largest $t_{\rm peak}/P_{\Delta\epsilon}$ (for $M_{\rm BH}=5\times 10^{7}$) and shortest one (for $M_{\rm BH}=10^{5}$) differ by a factor of $2-4$.  For $M_{\star}=1$, $t_{\rm peak}/P_{\Delta\epsilon}$ rises from 0.50 for $M_{\rm BH}=10^{5}$ to 0.55 for $M_{\rm BH}=10^{6}$, and 1.0 for $M_{\rm BH}=10^{7}$. These shifts are superimposed upon those created by the internal structure of the stars.

There are also finer-scale features that depend on black hole mass, such as the steepness of the initial rise and the shape of the peak.  $\dot{M}_{\rm fb}/\dot{M}_{0}$ for the $0.3\Msol$ star increases very sharply as a result of the sharp edge at the low-energy end of $dM/dE$, whereas $\dot{M}_{\rm fb}/\dot{M}_{0}$ for the $1\Msol$ and $3\Msol$ stars begins to rise sooner and approaches the peak more gradually due to the wider tails in their energy distributions.  In addition, the maximum in $\dot M_{\rm fb}/\dot M_0$ for $M_{\star}=1$ is rather flat and broad, particularly for larger $M_{\rm BH}$.

\begin{figure}
	\centering
	\includegraphics[width=8.8cm]{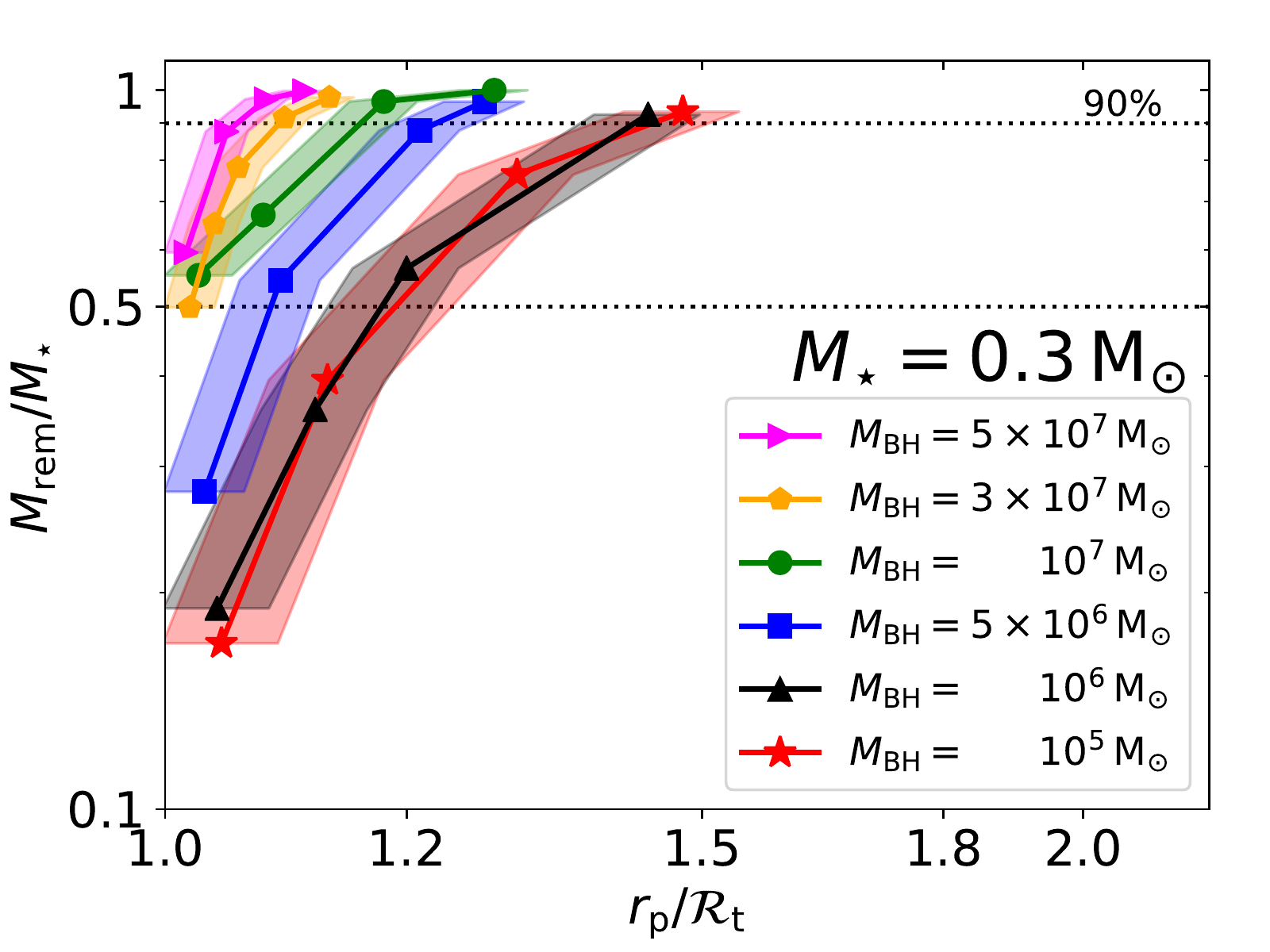}
	\includegraphics[width=8.8cm]{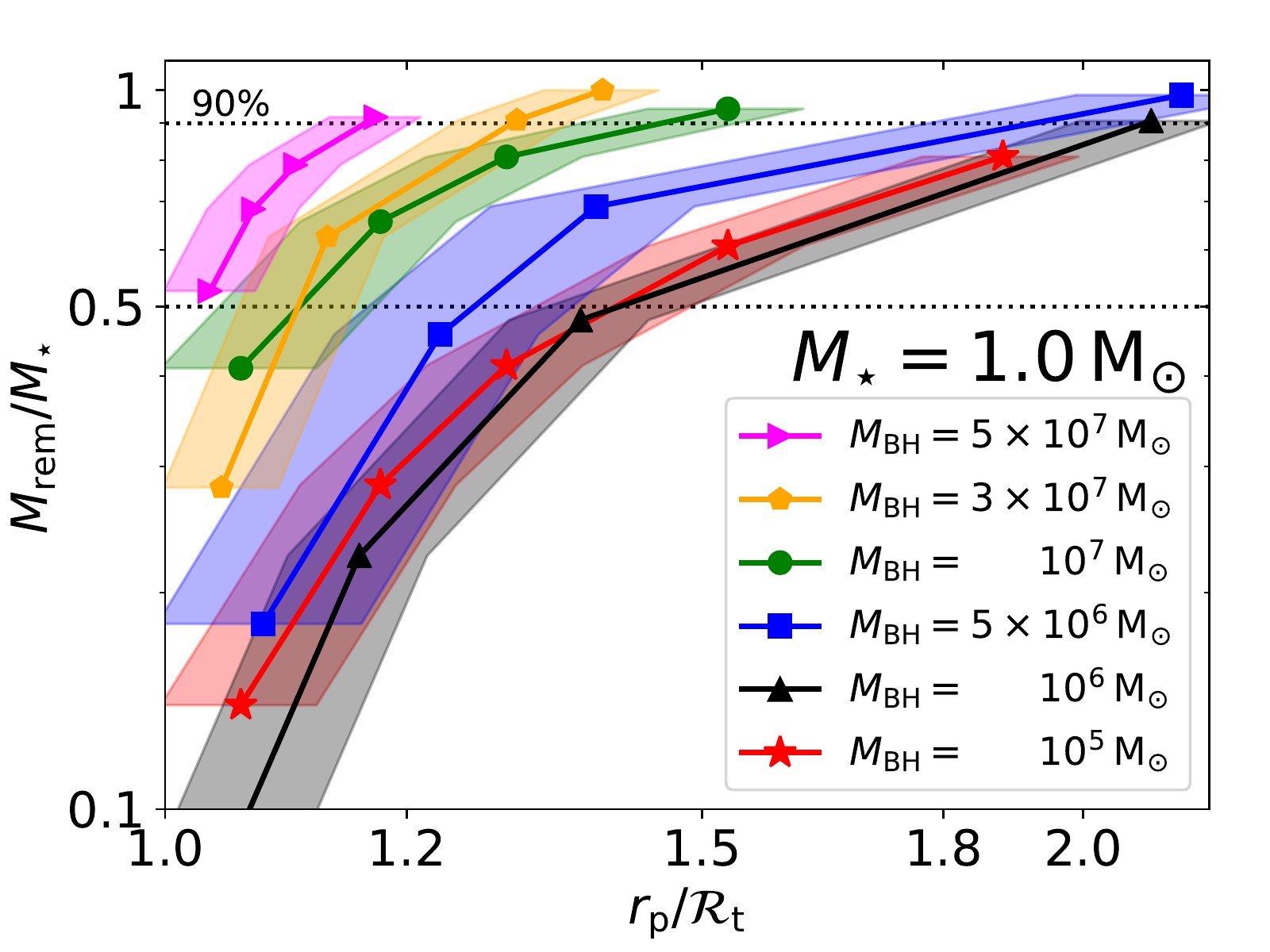}	
	\includegraphics[width=8.8cm]{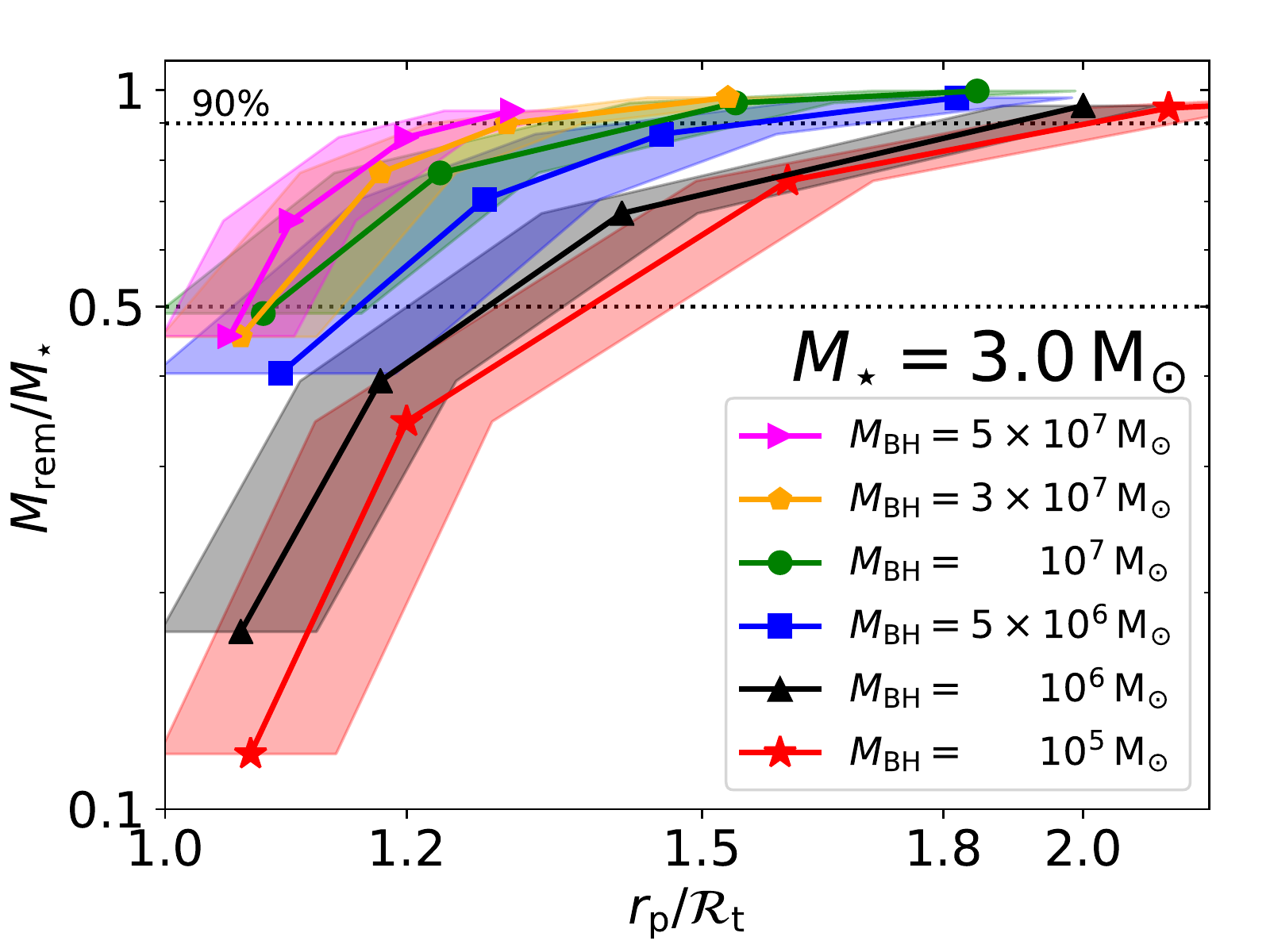}
	\caption{The fractional remnant mass $M_{\rm rem}/M_{\star}$ as a function of pericenter distance $r_{\rm p}$ normalized by the physical tidal radius $\physrad$ for $M_{\star}=0.3$ (\textit{top} panel), $1$ (\textit{middle} panel) and $3$ (\textit{bottom} panel). The 50\% and 90\% levels are marked by horizontal dotted lines. The shaded regions demarcate the ranges determined by the uncertainties of $\physrad$, filled with the same colors as the solid lines. }
	\label{fig:remnantmass}
\end{figure}

Because of cases like these, we do not define $t_{\rm peak}$ as the actual time when $\dot M_{\rm fb}\dot M_0$ reaches its absolute maximum, but rather as the time at which 5\% of $M_{\star}$ has returned to the black hole.  This time corresponds to the time of the absolute maximum when the peak is sharp, and the beginning of the maximum when the peak is relatively flat.  In addition, it is very close to the orbital period of matter with $E\simeq -\Delta E$, making it consistent with the traditional definition of the characteristic timescale of mass-return even though our $dM/dE$ distributions are not square waves.

Several previous efforts have been made to determine how relativistic dynamics alter fallback rates.
Using Newtonian and relativistic hydrodynamic simulations, \citet{Cheng+2014} studied the tidal encounters of a $1\Msol$ polytropic star with $\gamma=5/3$ with BHs of varying masses ($10^{5},~10^{6}$ and $10^{7}$).  The treatment of the star's self-gravity in their relativistic simulations is quite similar to ours: the self-gravity is calculated using a Newtonian Poisson solver in a frame comoving with the star and defined to be nearly Minkowski. The only difference is that they used Fermi normal coordinates to define this frame \citep[][]{ChengEvans2013} rather than a tetrad system as we did. The results from their relativistic simulations show a shift in $t_{\rm peak} / P_{\Delta\epsilon}$ with the same sign as ours, but significantly smaller amplitude: rather than a factor of 2--4 from $M_{\rm BH}=10^5$ to $M_{\rm BH}=10^7$, they found only a factor 1.1.

\citet{ServinKesden2017} also estimated $\dot{M}_{\rm peak}$ and $t_{\rm peak}$ using relativistic corrections to the energy width for $10^{5}\leq M_{\rm BH}\leq10^{7}$.
They found results qualitatively consistent with ours in that $\dot{M}_{\rm peak}$ decreases and $t_{\rm peak}$ increases. 
However, they found a significantly shallower slope for $t_{\rm peak}/P_{\Delta\epsilon}$ and $\dot M_{\rm fb}/\dot M_0$ between $M_{\rm BH}=10^5$ and $M_{\rm BH} = 10^7$ than we do. \citet{ServinKesden2017} predicted that $\dot{M}_{\rm peak}$ decreases by only 20\% from $M_{\rm BH}=10^{5}$ to $M_{\rm BH}=10^{7}$ whereas, over the same $M_{\rm BH}$ range, our calculations indicate that $\dot{M}_{\rm peak}$ decreases by a factor of 2.5.

\subsection{Partial disruption and the remnant mass}
\label{sub:remnantmass}

Stars are partially disrupted when $r_{\rm p} > \physrad$, but less than a few times $\physrad$ (\citetalias{Ryu3+2019}). 
Figure~\ref{fig:remnantmass} shows the ratio of the mass of the remnant to the initial stellar mass,  $M_{\rm rem}/M_{\star}$, as a function of $r_{\rm p}/\physrad$.
The mass of a remnant is defined as the mass enclosed in the computational domain when the mass settles to an asymptotic value.
The  fractional remnant masses for $M_{\rm BH}=10^{5}$ and $10^{6}$ are similar for a given $r_{\rm p}/\physrad$. However, for larger $M_{\rm BH}$, $M_{\rm rem}/M_{\star}$ at fixed $r_{\rm p}/\physrad$ grows.  In other words, for a fixed ratio of the pericenter to the physical tidal radius, stars are {\it better} able to hold onto their mass when the event is more realistic.  

\begin{figure}
	\centering
	\includegraphics[width=8.8cm]{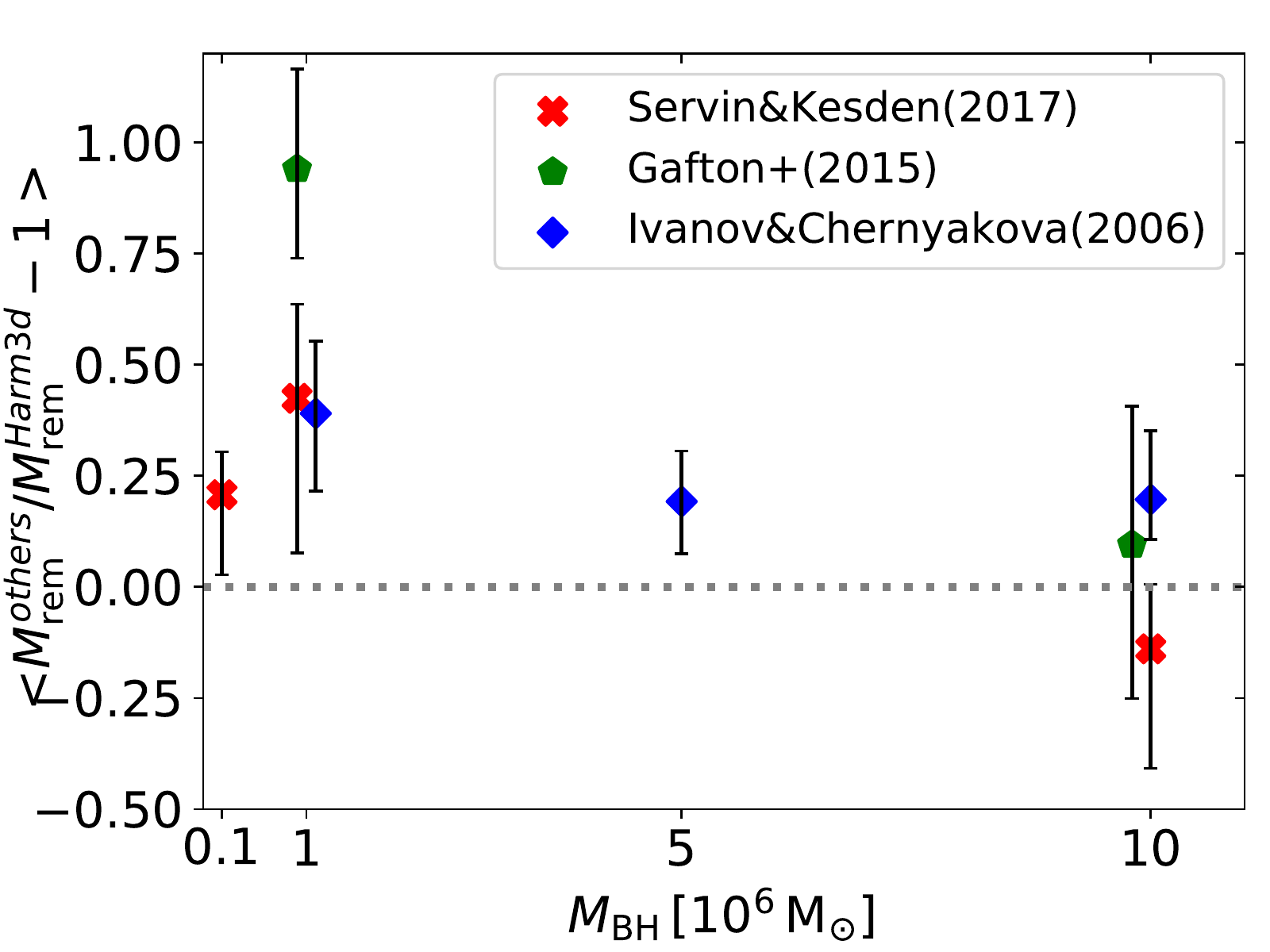}
	\caption{The average fractional difference between three other estimates of the remnant masses produced by partial disruptions ($M_{\rm rem}^{\rm others}$ as estimated by \citet{ServinKesden2017} [red crosses], \citet{Gafton+2015} [green pentagons] and \citet{IvanovChernyakova2006} [blue diamonds] ) and our simulations' estimates ($M_{\rm rem}^{\harm}$),  i.e., $\langle M_{\rm rem}^{\rm others}/M_{\rm rem}^{\harm}-1\rangle$. The error bars show the entire range of variation of the fractional differences over the span of $r_{\rm p}/\physrad$ shown in Figure~\ref{fig:remnantmass}. These ranges of variation are not standard deviations. For better readability, the symbols for $M_{\rm BH}=10^{6}$ from \citet{IvanovChernyakova2006} and for $M_{\rm BH}=10^{7}$ from \citet{Gafton+2015} are shifted horizontally by a small amount.}
	\label{fig:remnantmass_comparison}
\end{figure}

\citet{IvanovChernyakova2006}, \citet{Gafton+2015} and \citet{ServinKesden2017} also found that the remnant mass fraction for $1\Msol$ stars depends on $M_{\rm BH}$ in a fashion qualitatively similar to what we find, i.e., less mass is lost for higher $M_{\rm BH}$. For a more quantitative comparison, we used the curves shown in their papers to determine their expectation for $M_{\rm rem}/M_{\star}$ at values of $r_{\rm p}$ matching those used in our simulations. In Figure~\ref{fig:remnantmass_comparison}, we show the average fractional difference between $M_{\rm rem}/M_{\star}$ as found by the three studies (for $M_{\star}=1$) and the remnant mass fraction we determined. For almost the entire range of black hole mass considered, the values of $M_{\rm rem}/M_{\star}$ from \citet{IvanovChernyakova2006} and \citet{ServinKesden2017} are higher than ours by 20-60\%.
These rather small differences from ours are remarkable given the approximate methods used in these calculations.
Although the remnant mass fractions from \citet{Gafton+2015} are similar to ours for $M_{\rm BH}=10^{7}$, those for $M_{\rm BH}=10^{6}$ are higher by almost a factor of two.

\section{Implications}\label{sec:implication}

As our results illustrate, relativistic effects create $M_{\rm BH}$-dependence for all the principal properties of tidal disruptions: the physical tidal radius, the debris energy distribution, and the relation between orbital pericenter and remnant mass for partial disruptions.  These relativistic effects can produce quite noticeable departures from the Newtonian predictions for these physical quantities.

Relativistic effects also lead to significant changes in observable quantities.
Changes in the range of pericenters producing tidal disruptions translate directly into changes in event rates, particularly for galaxies in which the stellar angular momentum distribution is in the ``full loss-cone" limit.   Because the debris energy distribution determines the debris orbital period distribution, these changes alter the predicted fallback rate.  In this section we develop the consequences of these relativistic effects. 

This entire discussion is made simpler by our demonstration that the relativistic corrections to $\physrad$ and $\Delta E$ depend only very weakly on $M_{\star}$. The relativistic corrections to both $\physrad$ and $\Delta E$ can therefore be described by functions of $M_{\rm BH}$ wholly independent of $M_{\star}$. As we did in the previous three papers of this series, we refer to stars with $M_{\star}\leq0.5$ as ``low-mass'' stars and those with $M_{\star}\geq1$ as ``high-mass'' stars.

\subsection{Physical tidal radii}

The range in physical radii for main sequence stars of all masses at a single value of the black hole mass $M_{\rm BH} = 10^6$ is considerably narrower than would be predicted on the basis of $r_{\rm t}$ (Table~2 in \citetalias{Ryu2+2019} or the \textit{right} panel of Figure 3 in \citetalias{Ryu1+2019}).  From $M_{\star} = 0.15$ to $M_{\star}= 3$, the maximum pericenter at which a total disruption occurs has a range of only $\simeq 1.5$, whereas the range of $r_{\rm t}$ is $> 5$.  The reason for this narrowing is that the shape of the internal density profile as a function of $M_{\star}$ runs counter to the dependence of stellar radius on $M_{\star}$.

Because the relativistic corrections to $\physrad$ are nearly independent of $M_{\star}$, this range is almost preserved; in fact, the sense in which the relativistic corrections do depend mildly on $M_{\star}$ is such as to narrow the range even further (see Table~\ref{tab:psi}): at $M_{\rm BH} = 10^7$, it is only a factor of $\simeq 1.25$. Thus, for the great majority of main sequence stars, $\physrad$ is at most weakly dependent on $M_{\star}$ for any given $M_{\rm BH}$, no matter what that black hole mass is.

\begin{figure*}
	\centering
	\includegraphics[width=8.6cm]{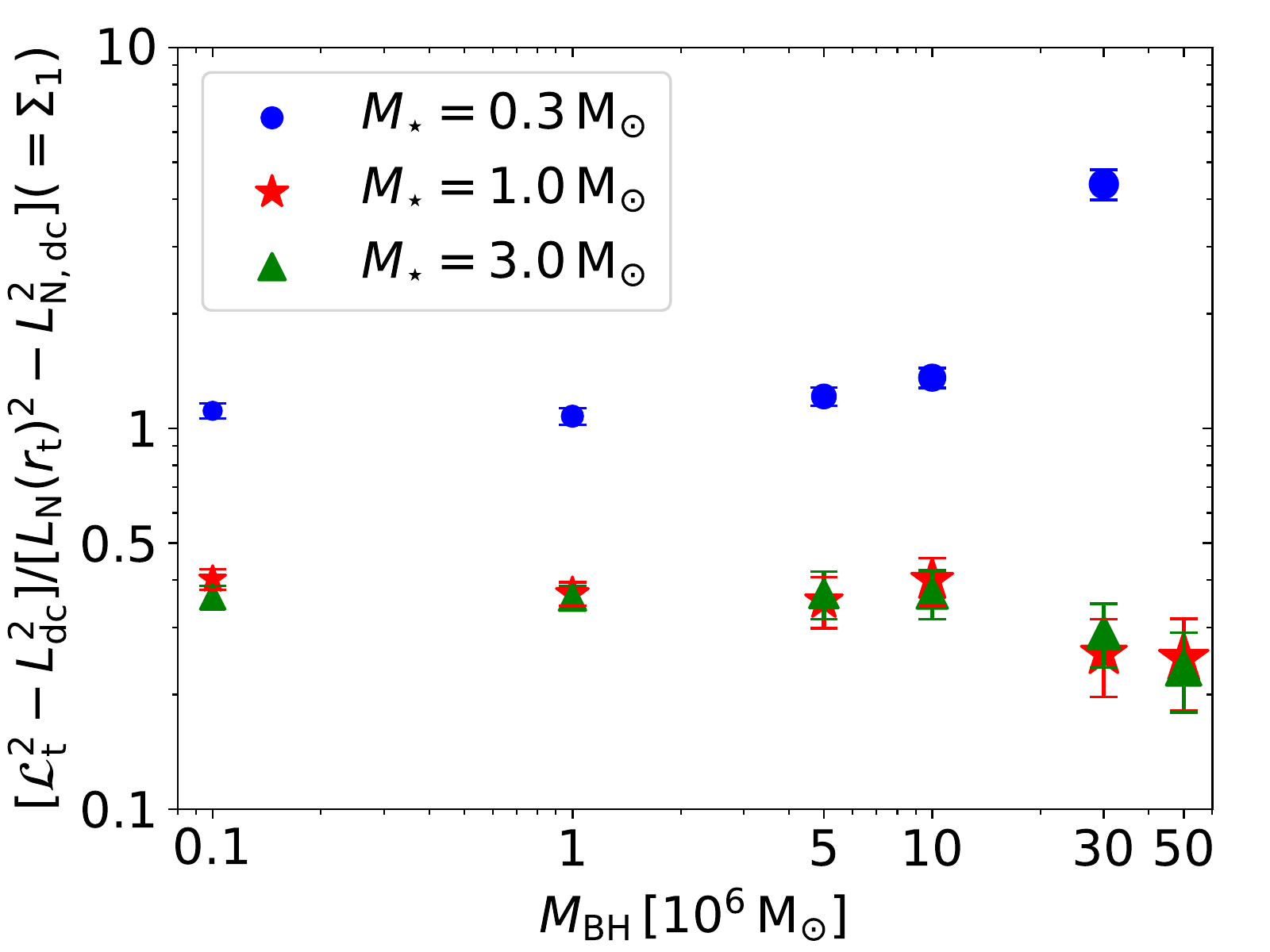}
		\includegraphics[width=8.6cm]{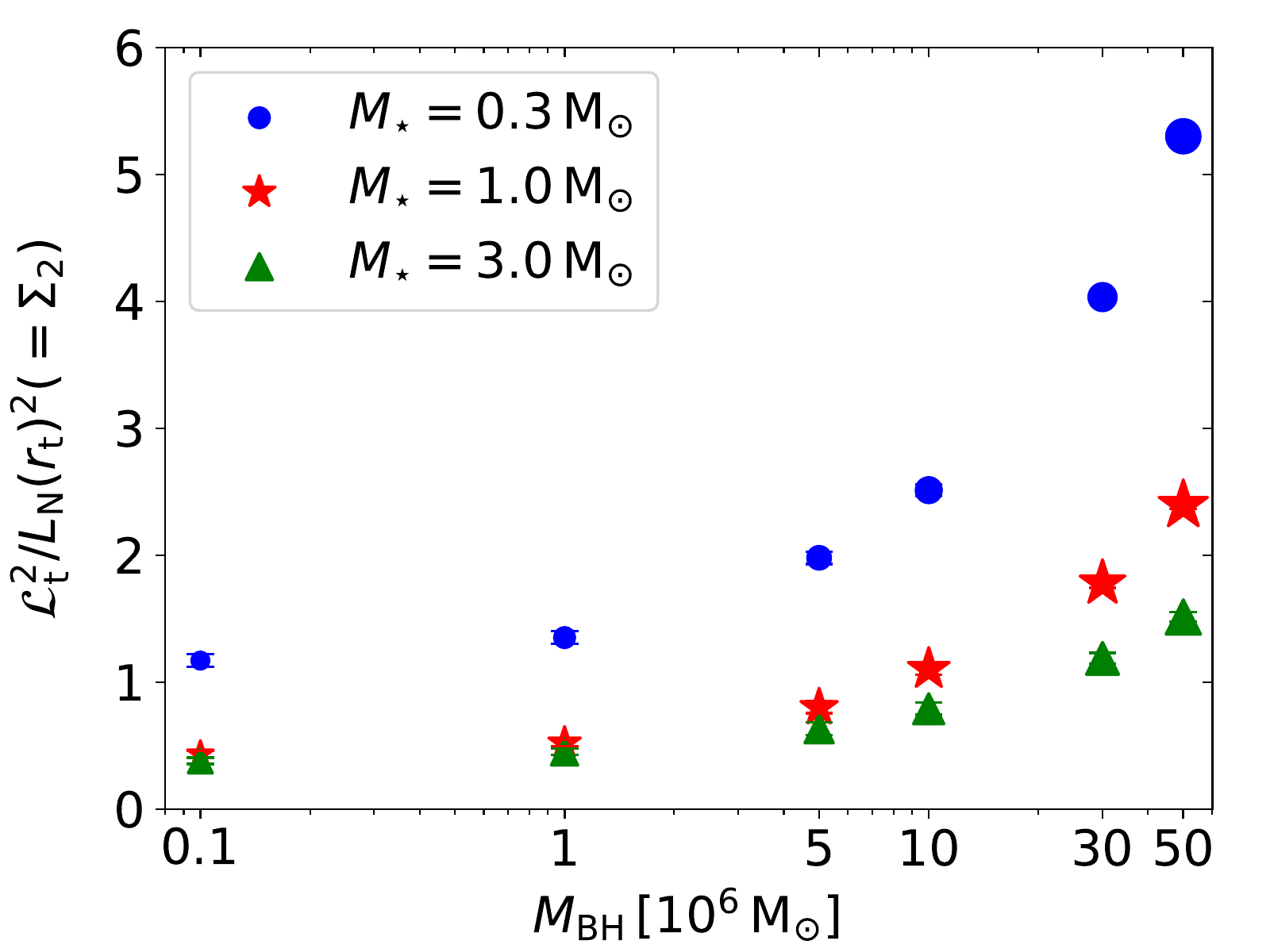}	
	\caption{(\textit{Left} panel) Ratio $\Sigma_1$, relevant to ``full loss-cone" event rates. (\textit{Right} panel) Ratio $\Sigma_2$, relevant to ``empty loss-cone" event rates. The error bars indicate the errors propagated from the uncertainties of $\physrad$.  }
	\label{fig:com_NT_GR}
\end{figure*}

\subsection{Relation between physical tidal radii and event rates}
\label{sub:TDErate}

The rate of TDEs depends on the specific angular momentum $L$ associated with an orbit whose pericenter is $\physrad$:
\begin{equation}\label{eq:L_GR}
L^2(r_{\rm p}=\mathcal{R}_{\rm t}) \equiv \mathcal{L}_{\rm t}^{2} = \frac{2 (\physrad/ \rg)^{2}}{\physrad/\rg - 2}.
\end{equation}
When the per-orbit root-mean-square change in $L$ is larger than  $\mathcal{L}_{\rm t}$ (the ``full loss-cone" or ``pinhole" regime), the stars' velocities (when far from the black hole) are distributed uniformly across the solid angle of the loss-cone.  It is then appropriate to speak of event ``cross sections". Because stars with $L < L_{\rm dc} (= 4r_{\rm g}c$ for parabolic orbits in Schwarzschild spacetime) plunge directly into the black hole without first being disrupted, the rate of total tidal disruptions is $\propto \mathcal{L}_{\rm t}^2 - L_{\rm dc}^2$ \citep{Kesden2012,Ryu1+2019}. 

On the other hand, when the rate at which a star's angular momentum changes is slow compared to the orbital frequency (the ``empty loss-cone" or ``diffusive" regime), the velocities of stars in the loss-cone are mostly directed very close to its edge. In this situation, the ``cross section" language is inappropriate because the distribution of impact parameters is not uniform. In this regime, the event rate depends logarithmically on $\mathcal{L}_{\rm t}$ \citep{LightmanShapiro1977,Merritt2013,Alexander2005} with a $\sim 10\%$ enhancement by occasional stronger encounters \citep{Weissbein2017}.  Direct capture is almost irrelevant in this regime until $M_{\rm BH}$ approaches the Hills mass.  Progression toward full disruption through the range of angular momenta larger than ${\cal L}_t$ is also interrupted by partial disruptions, which may lead to changes in the remnant's specific energy as well as its mass \citepalias{Ryu1+2019,Ryu3+2019}.

For these reasons, we focus here on how our calculations affect estimates of $\mathcal{L}_{\rm t}$, rather than their quantitative impact on actual event rates.

\subsubsection{Comparison between relativistic and estimated values of $\mathcal{L}_{\rm t}$}

For ``full loss-cone" angular momentum evolution, the rate of an event with $r_{\rm p} \leq \physrad$ is $\propto \Ltsq$, a quantity in which relativity alters the relation between $L$ and $\physrad$, and $\physrad$ itself differs from $\rtidal$ by effects both relativistic and derived from realistic stellar structure.  In addition, the actual rate of total disruptions is diminished by the rate at which direct capture, rather than tidal disruption, occurs.  On the other hand, in the ``empty loss-cone" regime (when one ignores the effects of partial disruptions), the rate is $\propto \ln(\mathcal{L}_{\rm t})$.

Consequently, to demonstrate how our predictions alter rates,  we examine two ratios:
\begin{align}
    \Sigma_{1} &= \frac{\mathcal{L}_{\rm t}^2 - L_{\rm dc}^2}{L_{\rm N}(r_{\rm t})^2-L_{\rm N,dc}^{2}},\\
    \Sigma_{2} &= \frac{\mathcal{L}_{\rm t}^2 }{L_{\rm N}(r_{\rm t})^2},
\end{align}
where the subscript N denotes the Newtonian functional relationship.
$\Sigma_1$ is the ratio between our predicted rate and the rate predicted by simple Newtonian estimates of disruption and direct capture; $\Sigma_2$ is the ratio between $\Ltsq$ and the square of the Newtonian angular momentum associated with the simple estimate.  The contrast between ``full loss-cone" event rates as we predict them and the simple estimate is given by the multiplicative factor $\Sigma_1$; the contrast between our predicted ``empty loss-cone" rates and those given by the traditional estimate is the additive factor $\ln \Sigma_2$.

The \textit{left} panel of Figure~\ref{fig:com_NT_GR} shows $\Sigma_{1}$ as a function of $M_{\rm BH}$. $\Sigma_1$  remains constant for $10^{5}<M_{\rm BH}<10^{6}$ because relativistic corrections remain relatively small for this range of $M_{\rm BH}$.  The departures from unity in $\Sigma_1$ in this range of $M_{\rm BH}$ reflect the corrections to the cross section due to our use of realistic internal stellar density profiles (for the low $M_{\rm BH}$ limit, $\Sigma_{1}\rightarrow \Psi$).  Above $M_{\rm BH} \approx 10^6$, $\Sigma_1$ for low-mass stars increases, while it falls for high-mass stars. This behavior is due to the competition between different relativistic effects, a competition that balances out differently depending on stellar structure.  Due to stronger tidal stress, $\physrad/r_{\rm t}$ increases with growing $M_{\rm BH}$, but the band of angular momentum outside $L_{\rm dc}$ and inside $\mathcal{L}_{\rm t}$ rapidly becomes narrower, approaching zero for $M_{\rm BH} > 5 \times 10^7$.  Stronger tidal stress plays the dominant role for $M_{\star}=0.3$, whereas the contribution from direct captures becomes more important for $M_{\star}=1$ and $3$.

The \textit{right} panel of Figure~\ref{fig:com_NT_GR} shows these comparisons for $\Sigma_2$, the parameter more relevant to the empty loss-cone limit.  Independent of stellar mass, this ratio increases with $M_{\rm BH}$ at an accelerating rate, reflecting the way in which stronger tidal stresses steeper relationship between $\mathcal{L}_{\rm t}^2$ and $r_{\rm p}$ when the orbit runs deep into the relativistic potential.  Unlike $\Sigma_1$, $\Sigma_2$ ignores losses due to direct capture.
$\Sigma_2$ grows by a factor of 3--5 from the Newtonian limit to $M_{\rm BH} = 5 \times 10^7$, depending on the stellar mass.

\subsubsection{Ratio of tidal disruption and direct capture cross sections in the full-loss cone regime}
\label{sub:TD_DC}

\begin{figure}
	\centering
	\includegraphics[width=8.9cm]{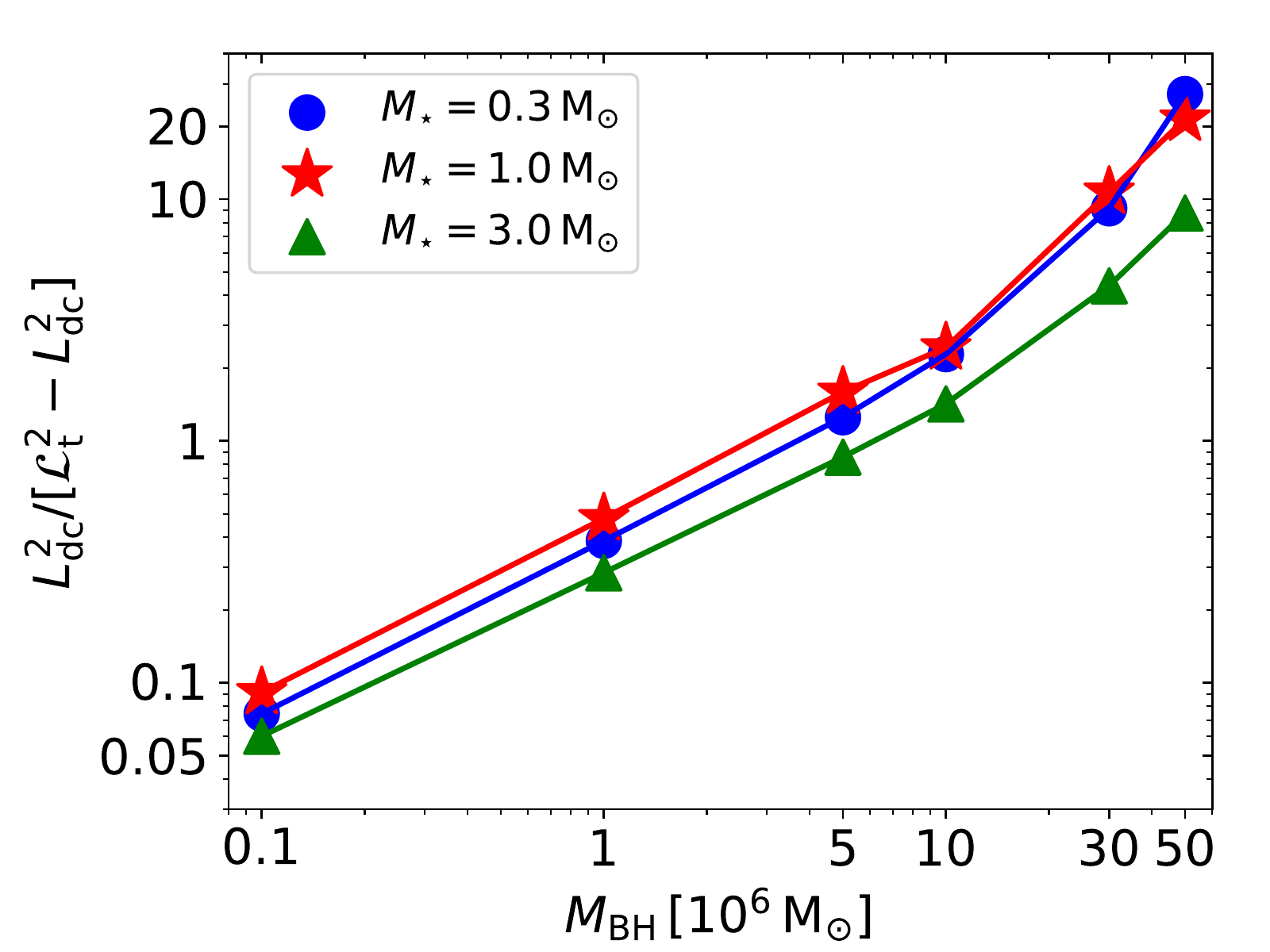}			
\caption{The ratio of direct capture cross section to full disruption cross section $L_{\rm dc}^{2}/[\mathcal{L}_{\rm t}^{2}-L_{\rm dc}^{2}]$ as a function of $M_{\rm BH}$. }
	\label{fig:com_TDE_DC}
\end{figure}

To illustrate how relativistic effects alter the outcome of tidal disruption events taking place in the full loss-cone context, Figure~\ref{fig:com_TDE_DC} shows the ratio of the cross sections for direct capture to those for full tidal disruptions, i.e., $L_{\rm dc}^{2}/[\mathcal{L}_{\rm t}^{2}-L_{\rm dc}^{2}]$ for the three stellar masses. 
This ratio increases from being rather small for low $M_{\rm BH} ($ $\lesssim 0.1$ for $M_{\rm BH} = 10^5$) to greater than unity for $M_{\rm BH} \gtrsim 5 \times 10^6$, although the precise value of the ratio depends weakly on $M_{\star}$.  It becomes $\gtrsim 10$ for $M_{\rm BH} \gtrsim 5 \times 10^7$.

\citet{Kesden2012} also estimated this ratio, but in a different framework.  His dynamical calculation also used relativistic tidal stresses and orbital dynamics, but he defilned $\mathcal{L}_{\rm t}$ by the condition that  the Newtonian surface gravity of a star with solar mass and radius match the magnitude of the eigenvalue for tidal stretch at the orbital pericenter; in other words, neither hydrodynamics nor the star's internal density profile played a role.  In addition, rather than present the cross section ratio, he presented the ratio of rates corresponding to a particular full loss-cone model.  This approach yielded $L_{\rm dc}^{2}/[\mathcal{L}_{\rm t}^{2}-L_{\rm dc}^{2}]$ at $M_{\rm BH} = 10^6$ $\sim 3-4 \times$ smaller than our value for $M_{\star}=1$, and a factor of 2 smaller for $M_{\rm BH} > 10^7$.  These quantitative contrasts may be due to both the stellar orbital population model used by \citet{Kesden2012} and the lack of hydrodynamics in his calculations.

\begin{figure}
	\centering
	\includegraphics[width=8.6cm]{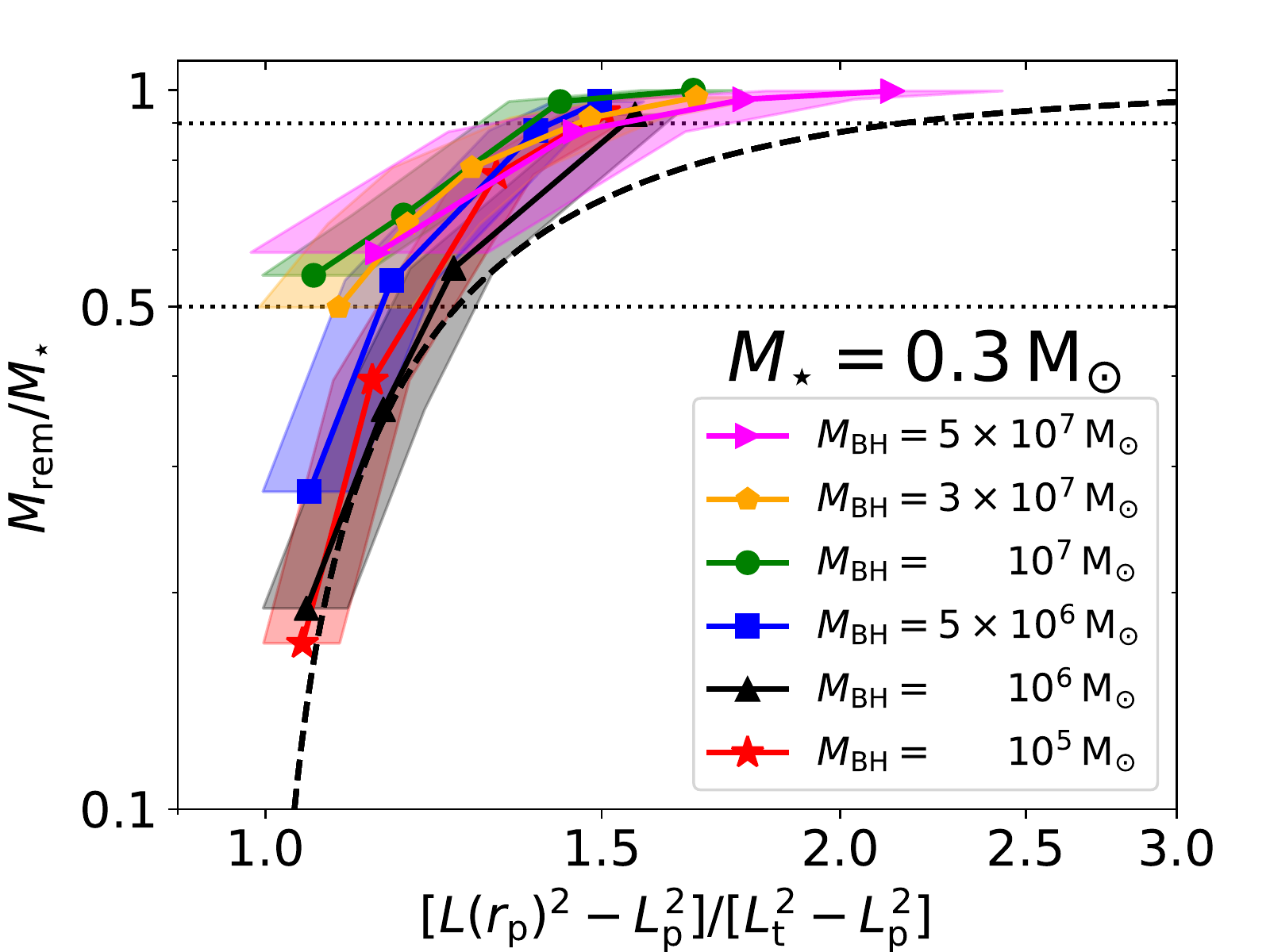}
	\includegraphics[width=8.6cm]{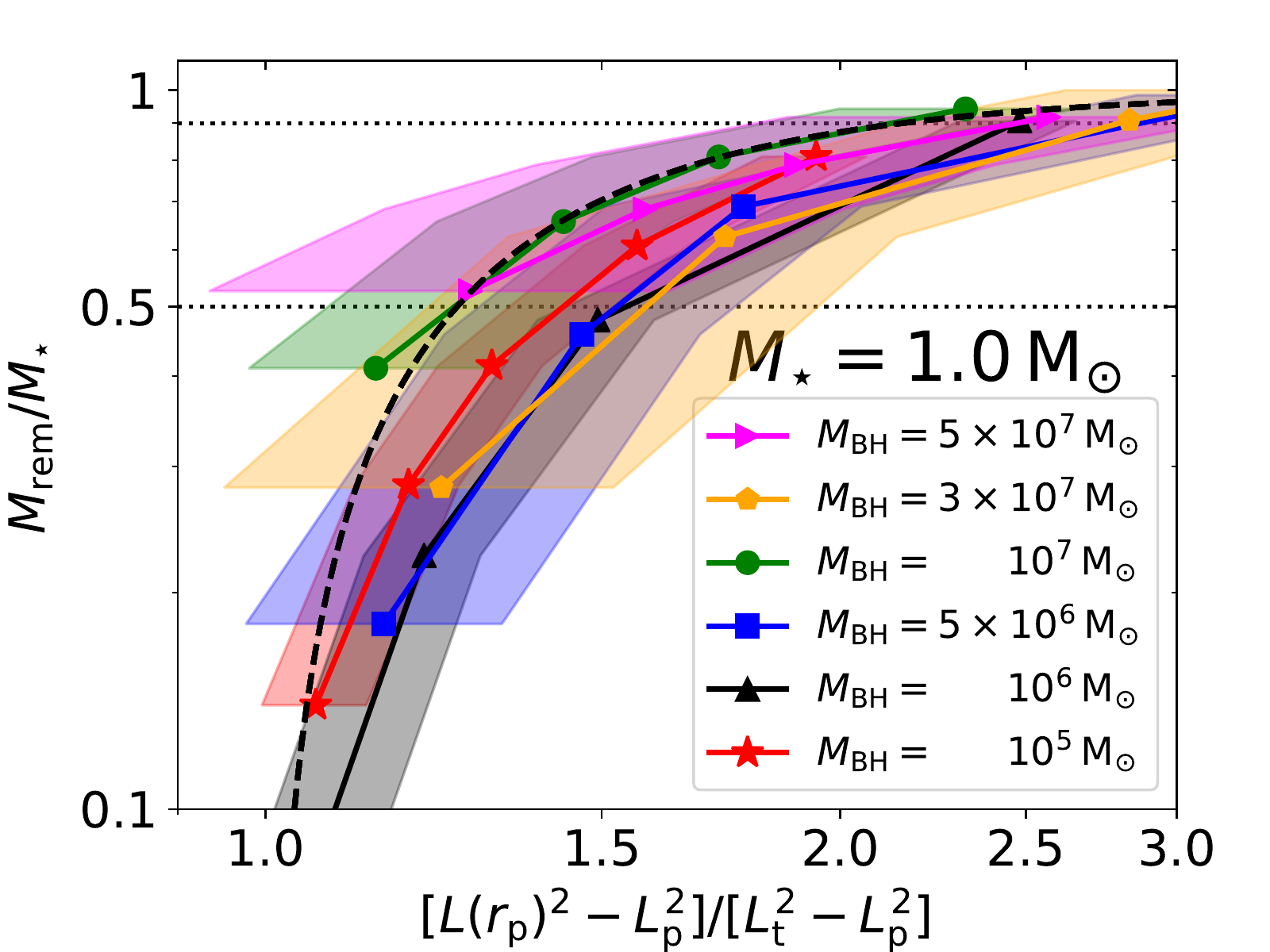}	
	\includegraphics[width=8.6cm]{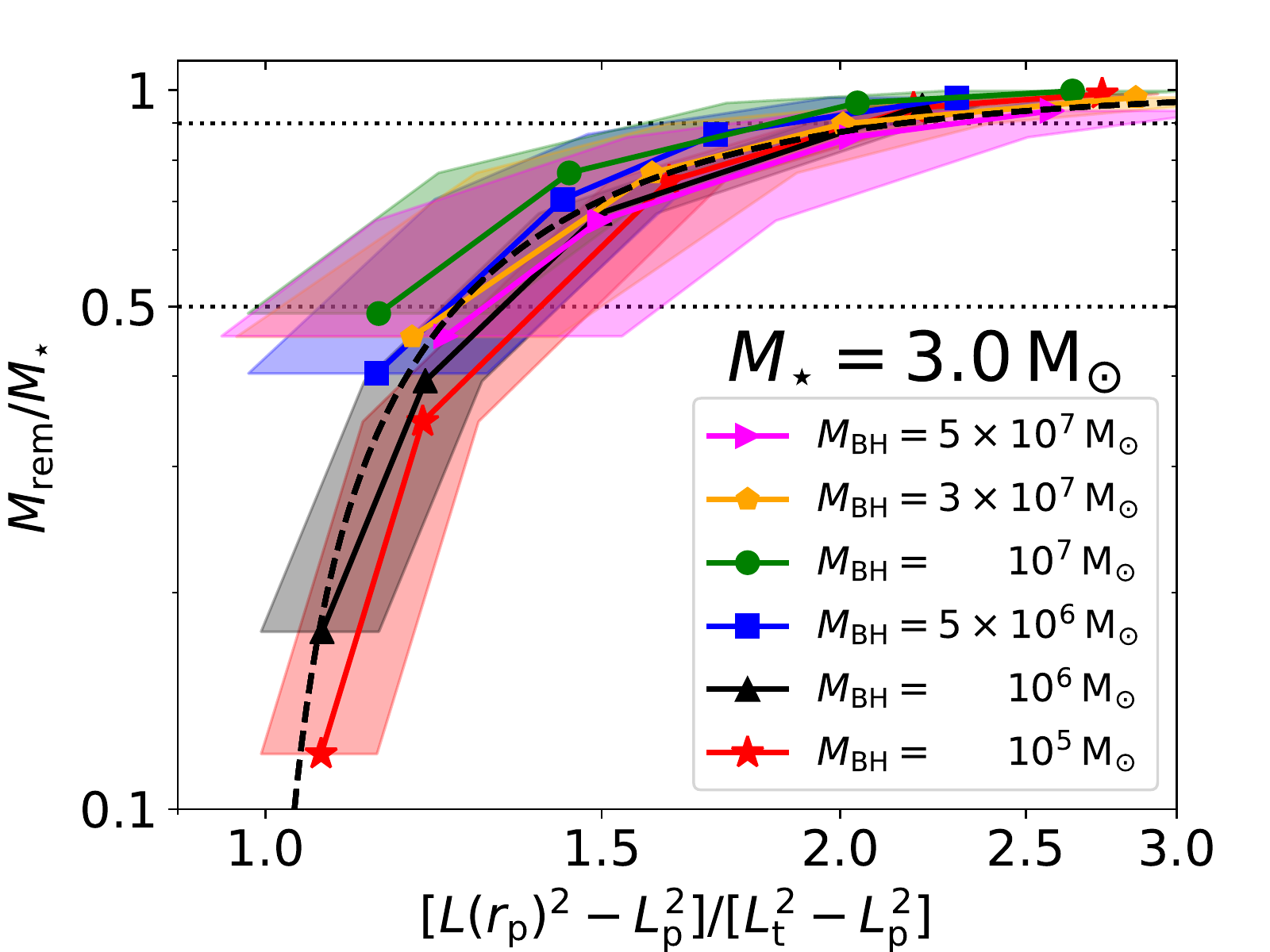}
	\caption{The fractional remnant mass $M_{\rm rem}/M_{\star}$ as a function of the ratio of the cross-section for full+partial to full disruptions for $M_{\star}=0.3$ (\textit{top} panel), $1$ (\textit{middle} panel) and $3$ (\textit{bottom} panel). The 50\% and 90\% levels are marked by horizontal dotted lines. The shaded regions delineate the uncertainties of the cross section ratio propagated from the uncertainties of $\physrad$, filled with the same colors as the solid lines. The black dashed lines in each panel depict the fit given in Equation~\ref{eq:remnant_fit}.}
	\label{fig:remnantmass_L}
\end{figure}

\subsubsection{Maximum black hole mass for tidal disruption}
\label{sub:max_mBH}

The replacement of tidal disruption with direct capture places a fundamental limit on the range of black hole masses relevant to TDEs.  Indeed, to the degree that we can be confident about this limit, it can be used to constrain the inference of $M_{\rm BH}$  in observed TDE events \citep[e.g.][]{Leloudas+2016}. However, the concept of ``maximum black hole mass" is necessarily somewhat fuzzy.  As shown by \citet{Kesden2012}, when the black hole has non-zero spin, the maximum mass depends on the black hole's spin parameter and the angle between the black hole's angular momentum and the star's orbital angular momentum.   More fundamentally, as was noted by \citet{Kesden2012} and can be seen in our study of the $M_{\rm BH}$-dependence of $\mathcal{L}_{\rm t}^2 - L_{\rm dc}^2$, even for masses a factor of several below the absolute maximum mass, the rate of tidal disruptions (when stellar angular momentum evolves rapidly, the ``full loss-cone" case) can be very strongly suppressed by the competition with direct capture. On the other hand, if the limit of slow stellar angular momentum evolution applies (the ``empty loss-cone" regime), a condition that might apply to spherical stellar distributions around high-mass black holes \citep{StoneMetzger2016}, direct capture is irrelevant until $M_{\rm BH}$ is large enough that $\Ltsq$ becomes very close to $L^2_{\rm dc}$.

In our special case of non-spinning black holes,  we define $M_{\rm BH,max}$ as the value of $M_{\rm BH}$ for which $L=L_{\rm dc}$, the angular momentum at which $\physrad = 4r_{\rm g}$ (note that the data presented in \citet{Kesden2012} indicate that $L_{\rm dc}$ is very weakly dependent on spin when the orientation of the orbital axis relative to the spin axis is averaged over solid angle). Because the smallest $\physrad/\rg$ in Table~\ref{tab:tidal_r_BH} is $\simeq 6-7$, we can not directly determine $M_{\rm BH, max}$ from the simulation results, but it is clear that $M_{\rm BH,max} > 5\times10^{7}$.
Note that our lower bound on $M_{\rm BH, max}$ is larger than some previous estimates, e.g., $M_{\rm BH,max} \simeq 2.5\times 10^{7}$ for a solar-type star suggested by \citet{ServinKesden2017}.  {On the other hand, we also find that the rate of direct capture becomes comparable to that of tidal disruption at a mass a factor $\sim 2$ {\it smaller}, so that the range of black hole masses in which the two rates compete is significantly broader than previously estimated.} The disagreement can probably be attributed to differences in method: \citet{ServinKesden2017} determined $M_{\rm BH,max}$ by defining $\mathcal{L}_{\rm t}$ in terms of a match between the Newtonian self-gravity and an eigenvalue of the relativistic tidal tensor, but adjusted with a parameter derived from the Newtonian calculations of \citet{Guillochon+2013} applied to polytropic stars.

\subsubsection{Ratio of partial to total disruption cross sections}\label{sec:full_partial}

Partial disruptions, by definition, involve stars {\it outside} the loss-cone.  For these stars, the cross section approach is appropriate.   It is then convenient to compare the rates for these events to the rates for total disruptions.  Just as for total disruptions, the cross section is $\propto L^2 = 2(r_{\rm p}/r_{\rm g})^{2}/(r_{\rm p}/r_{\rm g} - 2)$.

We show in Figure~\ref{fig:remnantmass_L} the remnant mass fraction $M_{\rm rem}/M_{\star}$ as a function of the ratio $[L(r_{\rm p})^{2}-L_{\rm dc}^{2}]/[\mathcal{L}_{\rm t}^{2}-L_{\rm dc}^{2}]$. This ratio compares the cross section for all events (full+partial) with pericenter up to $r_{\rm p}$ with the cross section for full disruptions; in the Newtonian limit, it reduces to $r_{\rm p}/{\cal R}_{\rm t}$.  The curves for different black hole masses coincide significantly more closely than the curves in Figure~\ref{fig:remnantmass}, where the same remnant mass fraction is plotted as a function of $r_{\rm p}/{\cal R}_{\rm t}$. 

Due to the near coincidence of the curves plotted in Figure~\ref{fig:remnantmass_L}, all of them can be described---to the same accuracy as our expression for the $M_{\rm BH}=10^6$ case---by a single curve, first presented in \citetalias{Ryu1+2019}:
\begin{align}\label{eq:remnant_fit}
\frac{M_{\rm rem}}{M_{\star}} = 1 - \left[\frac{L(r_{\rm p})^{2}-L_{\rm dc}^{2}}{\mathcal{L}_{\rm t}^{2}-L_{\rm dc}^{2}}\right]^{-3}.
\end{align}

\begin{figure}
	\centering
	\includegraphics[width=8.9cm]{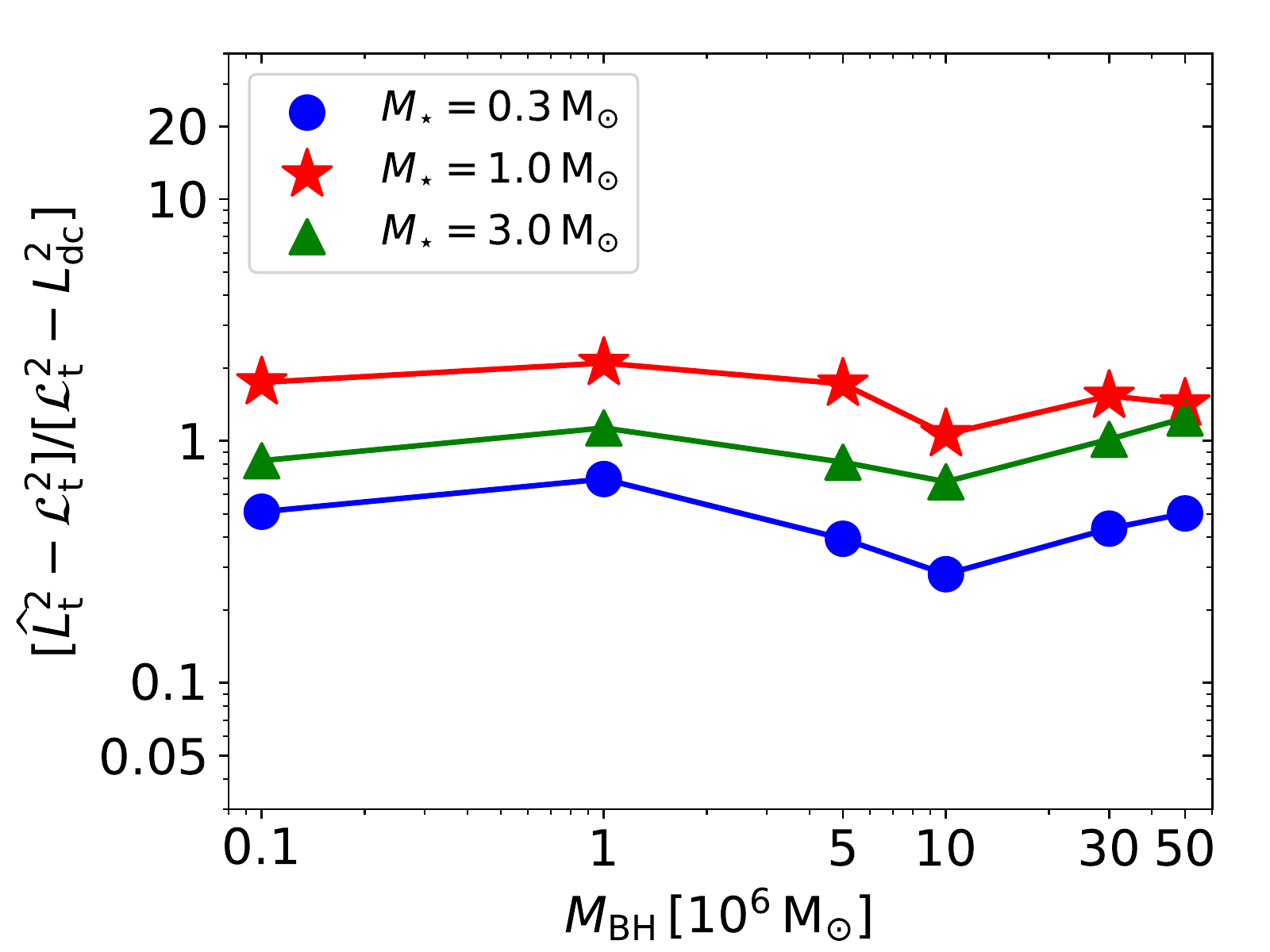}			
    \caption{ The ratio of the partial disruption to full disruption cross section $[\widehat{L}_{\rm t}^{2}-\mathcal{L}_{\rm t}^{2}]/[\mathcal{L}_{\rm t}^{2}-L_{\rm dc}^{2}]$, estimated from analytic fits to the remnant mass curves in Figure~\ref{fig:remnantmass}, as a function of $M_{\rm BH}$. }
	\label{fig:com_TDE_PF}
\end{figure}

The cross section ratio of all partial disruptions to all full disruption events is $[\widehat{L}_{\rm t}^{2}-\mathcal{L}_{\rm t}^{2}]/[\mathcal{L}_{\rm t}^{2}-L_{\rm dc}^{2}]$, which is depicted in Figure~\ref{fig:com_TDE_PF}. Here, $\widehat{R}_{\rm t}$ is the largest pericenter distance yielding partial disruptions. To use our data in order to measure $\widehat{R}_{\rm t}$,
we define it to be $r_{\rm p}$ for $M_{\rm rem}/M_{\star}=0.9$. We locate this point by linear interpolation between the two data points closest to $M_{\rm rem}/M_{\star}=0.9$.
Experimentation with other interpolation methods led to only slight changes in the results.

As is clear from Figure~\ref{fig:com_TDE_PF}, the ratio of the partial to full disruption cross section depends quite weakly on $M_{\rm BH}$, varying by less than a factor of two from the Newtonian limit to the highest black hole masses probed.  It does, however, depend somewhat on $M_{\star}$: it is $\approx 0.5$ for $M_{\star}=0.3$, $\approx 2$ for $M_{\star}=1$, and $\approx 1$ for $M_{\star}=3$. The weak $M_{\rm BH}$-dependence is because as $M_{\rm BH}$ increases, the full disruption cross section decreases due to direct capture events while the partial disruption cross section also declines owing to the decrease in $\widehat{R}_{\rm t}/\physrad$ (see Figure~\ref{fig:remnantmass}).

\section{Summary}
\label{sec:summary}

 This paper is the fourth in a series presenting the results of tidal disruption event simulations that, for the first time, combine general relativistic hydrodynamics, careful calculation of stellar self-gravity in a relativistic spacetime, and realistic main-sequence stellar structures for a wide range of stellar masses.  In this paper, we have focused on how properties of TDEs depend on black hole mass for non-spinning black holes; because the characteristic distance scales measured in gravitational units decrease with increasing $M_{\rm BH}$, studying TDEs at higher black hole mass means studying them in increasingly relativistic conditions.

 Although qualitative results have been obtained previously on some of the issues we consider
\citep{IvanovChernyakova2006,Kesden2012,ServinKesden2017,Gafton+2015,Tejeda+2017, Gafton2019},  our more powerful methods (see \citetalias{Ryu2+2019} for details) have enabled quantitative characterization---and therefore greater insight---about how TDE properties depend on $M_{\rm BH}$:

 $\bullet$ The dependence on $M_{\rm BH}$ of the maximum radius for total disruption $\physrad$ can be factored out from its weak dependence on $M_{\star}$.  We find that  for a fixed $M_{\star}$, the ratio of $\physrad$ to the classical estimator, $r_{\rm t}$, 
can be well approximated as $ \Psi_{\rm BH}(M_{\rm BH})\equiv \physrad/r_{\rm t} = 0.80 + 0.26~({M_{\rm BH}}/{10^{6}})^{0.5}$. This function can and should be used a simple correction factor for the Newtonian estimates. As
$M_{\rm BH}$ increases, this ratio steadily grows, increasing by a factor $\simeq 3$ from the Newtonian limit, $M_{\rm BH} = 10^5$ to the relativistic one, $M_{\rm BH} = 5 \times 10^7$.

$\bullet$ A direct corollary of the increase in $\physrad/r_{\rm t}$ is that the rate of events with pericenters $\leq \physrad$ increases, relative to a Newtonian estimate based upon $r_{\rm t}$, by a factor $\simeq 5$ from the Newtonian limit to $M_{\rm BH} = 5 \times 10^7$.  However, at the same time, the fraction of direct captures also increases, becoming a majority of these events for $M_{\rm BH} > 5 \times 10^6$.
Although our results are all calculated in Schwarzschild spacetime, they would change little in Kerr if averaged over orbital orientation because, as shown by \cite{Kesden2012}, the orientation-averaged angular momentum for direct capture in Kerr almost exactly coincides with Schwarzschild.  Our main-sequence structures and hydrodynamics permit us to calculate $\physrad$, and therefore the flare fraction.

$\bullet$
The Newtonian estimate $\Delta\epsilon$ for the width of the debris energy distribution is $\propto M_{\rm BH}^{1/3}$. However, the energy spread becomes narrower than this for higher SMBH masses: the  ratio of the actual energy width $\Delta E$ to  $\Delta\epsilon$ falls by a factor $\simeq 2$ from the Newtonian limit $M_{\rm BH} = 10^5$  to the relativistic regime, $M_{\rm BH} = 5 \times 10^7$. 
This lengthens the return time and reduces the return rate of the debris stream. 

$\bullet$ Despite all these strong dependences on $M_{\rm BH}$, the full loss-cone rates of partial disruptions and total disruptions remain approximately equal for all $M_{\star} \lesssim 3$ and across the entire range of $M_{\rm BH}$; the latter effect is due to the increasing fraction of direct captures as $M_{\rm BH}$ grows.  Still more surprisingly, the fraction of the star's incoming mass lost in a partial disruption can be reasonably approximated by a single function that depends {\it only} on the angular momentum of the star's orbit and ${\cal L}_t (M_{\rm BH})$, with almost no dependence on $M_{\star}$ or any separate function of $M_{\rm BH}$ (Equation~\ref{eq:remnant_fit}).

\section*{Acknowledgements}

This work was partially supported by NSF grant AST-1715032, Simons Foundation grant 559794 {and an advanced ERC grant TReX}. S.~C.~N. was supported by the grants NSF AST 1515982, NSF OAC 1515969, and NASA 17-TCAN17-0018, and an appointment to the NASA Postdoctoral Program at the Goddard Space Flight Center administrated by USRA through a contract with NASA.  This research project (or part of this research project) was conducted using computational resources (and/or scientific computing services) at the Maryland Advanced Research Computing Center (MARCC). The authors would like to thank Stony Brook Research Computing and Cyberinfrastructure, and the Institute for Advanced Computational
Science at Stony Brook University for access to the high-performance
SeaWulf computing system, which was made possible by a $\$1.4$M National Science Foundation grant (\#1531492).

\software{
matplotlib \citep{Hunter:2007}; \mesa~\citep{Paxton+2011}; 
\harm ~\citep{Noble+2009}.
}


\end{document}